\documentclass[12pt]{article}
\usepackage{geometry,enumerate,amsmath,amssymb}
\usepackage{fullpage}
\usepackage{graphicx}

\newcommand{\be}{\begin{equation}}
\newcommand{\ee}{\end{equation}}

\renewcommand{\hat}{\widehat}
\renewcommand{\tilde}{\widetilde}
\renewcommand{\epsilon}{\varepsilon}

\def\bz {\bar z}
\def\bZ {\bar Z}
\def\D {\Delta}
\def\S {\Sigma}
\def\Om {\Omega}
\def\p {\partial}
\def\tO{\tilde\Omega}
\def\M {{\cal M}_1}
\def\MN {{\cal M}_N}
\def\K {{\cal K}}
\def\tK {\tilde{\cal K}}
\def\half {\frac{1}{2}}
\def\pr {\partial}
\font\mybb=msbm10 at 11pt

\def\bb#1{\hbox{\mybb#1}}

\def\bR {\bb{R}}

\def\bC {\bb{C}}

\renewcommand{\theequation}{\arabic{section}.\arabic{equation}}
\newcommand{\news}{\setcounter{equation}{0}\quad}
\def\bea{\begin{eqnarray}}
\def\eea{\end{eqnarray}}
\begin{document}
\title{
\begin{flushright}\ \vskip -2cm {\small {DAMTP-2013-39 }}\end{flushright}
\vskip 10pt
\bf{Vortex Motion on Surfaces of Small Curvature}
\vskip 30pt}
\author{
Daniele Dorigoni, Maciej Dunajski and
Nicholas S. Manton\\[30pt]
{\em \normalsize
Department of Applied Mathematics and Theoretical Physics,}\\[0pt] 
{\em \normalsize University of Cambridge, Wilberforce Road, 
Cambridge CB3 0WA, U.K.}\\[10pt] 
{\small Email: D.Dorigoni@damtp.cam.ac.uk, 
\ M.Dunajski@damtp.cam.ac.uk, \ N.S.Manton@damtp.cam.ac.uk}\\[5pt]
}
\vskip 30pt
\date{14 August 2013}
\maketitle
\vskip 30pt
\begin{abstract}
We consider a single Abelian Higgs vortex on a surface $\S$ whose Gaussian curvature
$K$ is small relative to the size of the vortex, and analyse 
vortex motion by using geodesics on the moduli space of static solutions. 
The moduli space is $\S$ with a modified metric, and we propose that this metric
has a universal expansion, in terms of $K$ and its derivatives, around 
the initial metric on $\S$. Using an integral expression for the 
K\"ahler potential on the moduli space, we calculate the leading 
coefficients of this expansion numerically, and find some evidence for
their universality. The expansion agrees to 
first order with the metric resulting from the Ricci flow starting 
from the initial metric on $\S$, but differs at higher order. 
We compare the vortex motion with the motion of a point particle along geodesics of $\S$. 
Relative to a particle geodesic, the vortex experiences an additional force, which to 
leading order is proportional to the gradient of $K$. This 
force is analogous to the self-force on bodies of finite size that 
occurs in gravitational motion.
\end{abstract}
\newpage

\section{Bogomolny Vortices}\news

Solitons that satisfy Bogomolny equations experience no static
forces, and soliton motion is known to be well approximated by a geodesic 
motion in moduli space \cite{Ma1,book}. The metric on moduli space is
induced from the field kinetic energy and this dominates the dynamics 
because the potential energy is constant. For one soliton moving on a curved
base manifold, one may compare the geodesic motion in the 1-soliton 
moduli space with the geodesic motion of a test particle on the base 
manifold. They will differ because of the finite size of the soliton, 
and because of possible internal motion of the soliton, and it is 
interesting to study these effects. Here we calculate this 
difference for an Abelian Higgs vortex moving on a surface of small 
curvature. A vortex is a simple soliton, because it has a well defined 
location, and no internal motion. The 1-vortex moduli space, as a
manifold, is therefore the same as the base manifold, and only the
metrics differ.

In gravitational theory, it is one of the classic challenges to
describe the motion of a massive body of finite size in a given
gravitational background (see e.g. \cite{wald}), or the detailed interaction between two finite-size bodies \cite{Blanchet:2010cx}. Because a body's self-gravity 
locally dominates its background, it is difficult to compare the 
trajectory of the body with the geodesic representing a test particle. 
The trajectory also seems sensitive to the internal structure of the body.

Vortex motion provides a conceptually simpler set-up to consider this
issue. The mechanical properties of a vortex are determined by the
Abelian Higgs field theory and the background surface
geometry, but the background geometry is arbitrary and non-dynamical here, and not subject to an Einstein equation. 
A vortex has a precise centre, so its trajectory is unambiguous.
Our work thus gives some understanding of how the motion through curved
space of a body of small but finite size differs from the motion of 
a point-like test particle.

In the critically coupled Abelian Higgs theory, there is a moduli space
$\MN$ of static $N$-vortex solutions, which satisfy a coupled pair of 
Bogomolony equations. All these $N$-vortex solutions have the
same potential energy, so there are no static forces. The moduli space acquires a metric from the 
kinetic energy of the theory, and its geodesics model $N$-vortex motion. Mathematically, the metric is the restriction to
$\MN$ of the natural $L^2$ norm on the tangent space to the space of all field
configurations, with gauge freedom quotiented out. Samols found a
useful local expression for this metric on $\MN$ \cite{Sam}. The 
accuracy of geodesic motion on $\MN$, as an approximation to true 
$N$-vortex motion according to the field equations, was proved by 
Stuart \cite{Stu}.

The above discussion applies not just to $N$-vortex motion in $\bC$,
i.e. in the plane $\bR^2$, where the theory was originally defined,
but to $N$-vortex motion on any Riemann surface $\S$ that satisfies the Bradlow area inequality, which we recall below.
There is a modified Abelian Higgs theory, with Bogomolony 
equations and static $N$-vortex solutions, provided $\S$ has a 
metric compatible with the complex structure,
\be
g = \Om(z,\bz) dz d\bz \,,
\label{metric}
\ee 
where $z=x_1+ix_2$ is a local holomorphic coordinate. The function
$\Omega:\Sigma\rightarrow \bR^+$ is called the conformal factor of the
metric. The surface $\S$ should be metrically complete, and it may be 
compact and of finite area, or non-compact with boundaries at infinity. 

The fields of the theory on $\S$ are a complex scalar field, the 
Higgs field $\phi$, and an Abelian gauge potential $A_\mu$. They are defined 
on the 2+1 dimensional product space-time $\Sigma\times\bR$ with metric 
\be
ds^2=d{x_0}^2-g \,.
\ee
At the critical coupling the Lagrangian takes the form
\be
\label{lagrangian}
L=\frac{i}{2} \int_\Sigma 
dz\wedge d\bar{z}\,\Omega(z, \bar{z}) 
\left[-\frac{1}{4}F_{\mu\nu}F^{\mu\nu}
+\frac{1}{2} D_{\mu}\phi\overline{D^{\mu}\phi}
-\frac{1}{8}\left(\frac{1}{\tau}-|\phi|^2\right)^2\right],
\ee
where $F_{\mu\nu}=\pr_\mu A_\nu - \pr_\nu A_\mu$, and the Bradlow 
parameter $\tau $ is a positive constant. If $\S$ has a boundary, 
then $|\phi|^2 = \frac{1}{\tau}$ there.

The Lagrangian can be split into kinetic and potential terms
$L=T-V$, where $T$ is the part of $L$ containing derivatives w.r.t. $x_0$.
Completing the square in the potential energy $V$ one finds that, for 
a given magnetic flux, the total energy $E=T+V$ is minimal if the 
fields $(\phi, A_\mu)$ are static, with $A_0 = 0$, and satisfy the 
Bogomolny equations
\be
{D}_{\bar{z}}\phi=0 \,, \quad F_{12}=\frac{\Omega}{2}\Big(\frac{1}{\tau}
-|\phi|^2\Big) \,.
\ee
Here ${D}_{\bar{z}}=\p_{\bar{z}}-iA_{\bar{z}}$ is the
anti-holomorphic part of the $U(1)$ covariant derivative.  
The vortex number $N$ is the number of zeros of $\phi$, counted 
with multiplicity. The Bogomolny equations (with our choice of signs) 
only admit vortices of positive multiplicity, and solutions
generically consist of $N$ single vortices at distinct locations. The
potential energy of $N$ vortices is $N\pi/ \tau$, independently of 
their locations. Their total magnetic flux is $2\pi N$.

If the surface $\S$ is compact, its area is 
\be
A =  \frac{i}{2}\int_{\S} dz \wedge d\bz \, \Om(z,\bz)\,.
\ee
Bradlow \cite{Bra} showed, by integrating the second Bogomolny 
equation, that $N$-vortex solutions exist on $\S$ only if 
$A > 4\pi N\tau$. This can be interpreted as saying that the area of 
a vortex is $4\pi\tau$ and the total area occupied by vortices must 
be less than the area of the surface. We shall mostly be concerned
with compact surfaces of very large area, $A \gg 4\pi\tau$, or non-compact 
surfaces of infinite area, on which solutions exist for all $N$.
It is sometimes convenient to rescale $\tau$ to unity. One must then, 
at the same time, rescale the size of the surface $\S$ to retain
similar physics. 

Using the parametrisation $|\phi|^2 = \frac{1}{\tau} e^{h}$, and 
eliminating the gauge potential from the first Bogomolny equation one
finds that the second Bogomolny equation reduces 
to the gauge invariant Taubes equation \cite{Taubes}
\be
\Delta h + \frac{1}{\tau}\left(1-e^h\right) 
= \frac{4 \pi}{\Omega} \sum_{i=1}^N \delta^2(z-Z_i)
\label{Taubeseq}
\ee
where $Z_i$ are the vortex locations and $\D=4\Omega^{-1}\p_z\p_{\bar{z}}$ is 
the covariant Laplacian on $\S$. This is the fundamental tool for 
studying vortices on $\S$ and their moduli space. The 
Bogomolny equations do not directly give the delta functions, but they 
occur because of the logarithmic singularity of $h$ wherever $|\phi| = 0$.

As a manifold, the $N$-vortex moduli space $\MN$ is the 
symmetrised $N$-th power of $\S$,
\be
\MN = \S^N/S_N \,,
\ee
where $S_N$ is the permutation group. This is because an $N$-vortex
solution is completely determined by the $N$ unordered zeros of 
$\phi$, whose locations $\{Z_i: i=1,2,\dots,N\}$ are anywhere on 
$\S$ (and can coincide). $\MN$ is a smooth complex manifold, whose
natural coordinates are the symmetric polynomials in $\{ Z_i \}$.

The metric on $\MN$ is not explicitly known, in general. However,
using the Taubes equation, and its linearisation when the 
vortex locations are infinitesimally varied, Samols showed that 
the metric could be expressed in a localised form using $\{Z_i\}$ as
holomorphic coordinates \cite{Sam}. The metric coefficients depend on 
local data obtained from $h$ in the neighbourhoods of these vortex 
centres. From Samols' formula it can be deduced that the
metric on $\MN$ is K\"ahler. It is also smooth, even where vortex centres coincide. The cohomology class of the K\"ahler
2-form on $\MN$ can be determined \cite{MN}, so if $\S$ is 
compact and $A$ is finite, the volume of $\MN$ can be found. It 
depends on $N,A$ and the genus ${\tt g}$ of $\S$. 

Subsequent to Samols' work, Chen and Manton \cite{CM} found an
expression for the K\"ahler potential on $\MN$ in the case
of $N$ vortices on $\bC$, with its flat metric. This K\"ahler potential 
involves an integral over $\bC$ of an elementary function of $h$. 
However, the integral has logarithmic divergences at each point $Z_i$, 
and these need to be regularised, again using local data from the 
neighbourhoods of these points. Remarkably, the regularised integral can
be interpreted as the action whose Euler-Lagrange equation reproduces
the Taubes equation. It is the on-shell action, by which we mean the action 
evaluated on a solution of the Taubes equation, that is the main 
contribution to the K\"ahler potential. This on-shell action is not 
the standard potential energy of vortices in the Abelian Higgs theory, and is a non-trivial function of the vortex 
locations. In the Appendix we will re-derive the K\"ahler potential 
on $\M$, obtaining an expression that is valid for any curved surface $\S$. 

\section{The 1-Vortex Moduli Space}\news

The simplest example of moduli space geometry is for one vortex on 
the flat plane, $\bC$. The moduli space $\M$ is $\bC$, with the same 
flat metric. (This means a factor $\pi/ \tau$, representing the 
mass of the vortex, which occurs in the kinetic energy, has been scaled out.) 
Geodesic motion is straight line motion at constant speed. This 
accurately describes the non-relativistic limit of the exact solution
where a static vortex is Lorentz boosted.

In the case that $\S$ is curved, even the 1-vortex moduli space $\M$ is
geometrically interesting \cite{ManRic}. A 1-vortex has a single Higgs zero at
an arbitrary location $z=Z$, and is the unique solution of the Taubes equation
\be
\Delta h + \frac{1}{\tau}\left(1-e^h\right)  = \frac{4 \pi}{\Omega}
\delta^2(z-Z) \,.
\label{Taubes1}
\ee
The moduli space is therefore $\M = \S$, with $Z$ as coordinate, and 
Samols' analysis implies that the metric on $\M$ has the form
\be
\label{tgg}
\tilde{g} = \tO(Z,\bZ;\tau) dZ d\bZ \,.
\ee
The change of notation, from $z$ to $Z$, is slight,
and it is convenient to think of the metric on $\M$ as a modified 
version of the original metric $g$ on $\S$. Notice that the complex 
structure is unchanged in going from $\S$ to $\M$; just the conformal 
factor changes, from $\Om$ to $\tO$. In this 1-vortex case, Samols' 
formula simplifies to
\be
\tO(Z,\bZ;\tau) = \Om(Z,\bZ) + 2\tau\frac{\pr b}{\pr Z} \,,
\label{Sam1}
\ee
where $\half b(Z,\bZ)$ is the coefficient of $\bz-\bZ$ in the
expansion of $h$ around the vortex location $Z$,
\bea
h(z,\bz) &=& 2\log|z-Z| + a(Z,\bZ) 
+ \half {\bar b}(Z,\bZ) (z-Z) + \half b(Z,\bZ) (\bz-\bZ) \nonumber \\
&&\label{hexpans}
+ {\bar c}(Z,\bZ)(z-Z)^2 + d(Z,\bZ)(z-Z)(\bz-\bZ) + c(Z,\bZ)(\bz-\bZ)^2
+ \cdots \,.
\eea
Apart from the leading logarithmic term, this expansion is a Taylor
series in $z-Z$ and its conjugate. The Taubes equation (\ref{Taubes1})
requires that $d(Z,\bZ) = - \Om(Z,\bZ) / 4\tau$, but the other
coefficients shown here are not determined purely locally, but only 
from the complete 1-vortex solution. They also depend on $\tau$.

The function $b(Z,\bZ)$ is not explicitly known, except on some
especially symmetric surfaces, e.g. a round sphere, so
$\tO$ is not known either. Despite lacking explicit knowledge of the 
function $b(Z,\bZ)$, one can show that if $\S$ is compact, and of
genus ${\tt g}$, then the total area of $\M$ is 
${\cal A}_1 = A - 4\pi\tau(1-{\tt g})$ \cite{MN}. This result is a consequence
of being able to integrate $\frac{\pr b}{\pr Z}$. In detail, it relies on $h$
being a globally defined function of $z$ and $Z$ (and their
conjugates), with a singularity of type $2\log|z-Z|$, which implies a
particular type of transformation for $b$ under holomorphic changes 
of the local coordinate $z$.

The principal aim of this paper is to gain an understanding of the
conformal factor $\tO$ in the case that $\S$ has small curvature. We 
expect $\tO$ to differ rather little from $\Om$. 

To be precise about what is small in our approach, consider a metric 
$g_0 = \Om_0(z,\bz) dz d\bz$ on $\S$ that has Gaussian curvature $K_0$
and derivatives of $K_0$ all of order 1, and now assume that 
$\Om = L^2 \Om_0$, where $L$ is a large constant scale factor. Lengths 
get rescaled by $L$ and areas by $L^2$. The Gaussian curvature $K$ of $\S$ is
\begin{eqnarray}
K &=& -\half \D(\log \Om) \ 
\label{Gauss}\\
 &=& -\frac{1}{2L^2} \D_0(\log \Om_0 + 2\log L) = \frac{1}{L^2} K_0 \,.
\end{eqnarray}
So $K$ is small, of order $\frac{1}{L^2}$. Similarly $K^2$ and 
$\D K$ are of order $\frac{1}{L^4}$. Each factor of $K$ and each
application of $\D$ brings in a further factor of $\frac{1}{L^2}$.

Now, a vortex is a smooth localised solution of the Bogomolny
equations, centred at $Z$, whose characteristic area is of order $\tau$. It
is sensitive only to the local aspects of the background metric and
curvature near $Z$ if the surface is large and the curvature small. 
Long-range effects due to the curvature, including topological effects, 
are exponentially small, and we neglect them. These remarks also apply 
to the conformal factor of the moduli space, $\tO$. We propose
that $\tO$ has an asymptotic expansion in $\frac{\tau}{L^2}$ if expressed
in terms of the conformal factor $L^2 \Om_0$ and its curvature. If we
work directly with $\Om$ and its curvature $K$, then $\tau$
occurs explicitly, but the inverse powers of $L$ occur implicitly.

The expansion we propose (provisionally) is
\be
\tO(Z,\bZ;\tau) = \Om(Z,\bZ)(1 + \alpha_0 \tau K + \beta_0 \tau^2 K^2 
+ \gamma_0 \tau^2 \D K + \cdots) \,,
\label{tOexpans0}
\ee
where $\alpha_0, \beta_0, \gamma_0, \dots$ are universal constants of
order 1, independent of the function $\Om$. $K$ and its derivatives are
all evaluated at $Z$. Pulling out the overall factor $\Om$ is
dimensionally right, and ensures that all the terms inside the bracket
are invariant under holomorphic coordinate transformations. $\tau K$
is also dimensionless, and $\D$ has the same dimension as $K$. We have 
explicitly shown all the terms up to order $\frac{\tau^2}{L^4}$ that
can occur. This expansion is a refinement 
of what was proposed and studied in \cite{ManRic}.

We in fact know more about the nature of this expansion, because of
our precise knowledge of the area of $\M$ when $\S$ is
compact \cite{MN}. Because ${\cal A}_1 = A - 4\pi\tau(1-{\tt g})$, 
independently of $L$, we know that $\alpha_0 = -1$. This follows from 
the Gauss--Bonnet formula 
\be
 \frac{i}{2} \int_{\S} dZ \wedge d\bZ\,\Om\,K  = 4\pi(1-{\tt g}) \,.
\ee
Moreover, all subsequent terms in the expansion, involving higher
powers of $K$ and the operator $\D$, must integrate to zero. For this
to occur universally, the only terms allowed inside the bracket 
must be of the form of $\D$ applied to some further globally defined 
scalar expressions (as $\Om\D=4\p_Z\p_{\bar Z}$ is a total derivative, 
and a vortex is a scalar soliton, with no internal degrees of freedom). 
Therefore $\beta_0 = 0$, but $\gamma_0$ can be non-zero. We may
therefore rewrite the expansion (\ref{tOexpans0}) in its final form 
(and with new coefficients) as
\be
\tO(Z,\bZ;\tau) = \Om(Z,\bZ)(1 - \tau K + \alpha \tau^2 \D K 
+ \beta \tau^3 \D K^2 + \gamma \tau^3 \D^2 K + \cdots) \,,
\label{tOexpans}
\ee
where we have written out all terms up to order $\frac{\tau^3}{L^6}$. This holds for both compact and non-compact $\S$.

When $\Sigma$ is non-compact, simply-connected and asymptotically planar, a slight generalization of the Gauss--Bonnet formula yields
\be
 \frac{i}{2} \int_{\S} dZ \wedge d\bZ\,\Om\,K  = 0 \label{GauBonNC} \,.
\ee
This equation combined with (\ref{tOexpans}) tells us that, whether the area deficit between $\S$ and the flat plane, obtained by integrating $\Om-\Om_{flat}$, is finite or infinite, the deficit area between $\M$ and $\S$ is always zero, as
\be
 \frac{i}{2} \int_{\S} dZ \wedge d\bZ\, (\tO-\Om)  =  \frac{i}{2} \int_{\S} dZ \wedge d\bZ\, \left[- \tau\,\Om\,K  + \partial_Z \partial_{\bZ} (...)\right]=0\,.\label{NonCompact}
\ee

Ideally, we would now calculate the coefficients $\alpha, \beta,
\gamma, \dots$ using a general argument. However this would require
constructing a 1-vortex solution in a completely general background of
small curvature. Such a solution is close to the solution in flat space,
and it may be possible to construct it iteratively using a Green's
function. However, neither the flat space solution nor the relevant
Green's function are known in closed form, so we have not been able to
pursue this approach.

Instead, we have exploited the proposed universality of the
coefficients, and have calculated the first few of them -- $\alpha, \beta,
\gamma$ -- using a few carefully selected examples of conformal factors 
$\Om$. These conformal factors have more than just three
parameters, and for these we have verified the universality of the
coefficients. We have set $\tau = 1$ but have included 
$L$ as a parameter and taken the limit $L \to \infty$ to extract the
coefficients. In this way we avoid contamination by the neglected
higher order terms.

Our method involves solving Taubes' equation numerically. The background 
metric $g = \Om dzd\bz$ is axially symmetric around the origin $z=0$ in 
all cases, and is
asymptotic to the flat planar metric of $\bC$ as $|z| \to \infty$. So the
curvature is concentrated in a neighbourhood of the origin. If the
vortex is also located at the origin, then Taubes' equation reduces to
an ODE, which is straightforward to solve. From the solution we can
extract $a$ and $b$, the leading coefficients occurring in the expansion
(\ref{hexpans}) of $h$. However, this method now runs into difficulty. The
conformal factor $\tO$ involves not $b$ (which actually vanishes at the origin 
if there is axial symmetry), but $\frac{\pr b}{\pr Z}$, and to find this
derivative we would need to relocate the vortex away from the
origin. Accurately solving Taubes' equation, with an off-centre 
delta function, would not be numerically simple.

We have found a way round this. Rather than trying to find
$\tO$, we instead calculate its K\"ahler potential $\tK$. The general 
relation between $\tK$ and the conformal factor $\tO$ is 
\be
\tO = \pr_Z \pr_{\bZ} \tK \, .
\ee
Let $\K$ be the K\"ahler potential of the background $\Om$, so 
$\Om = \pr_Z \pr_{\bZ} \K$. The expansion (\ref{tOexpans}) can be integrated to
give
\be
\tK(Z,\bZ;\tau) = \K(Z,\bZ) + 2\tau\log\Om + 4\alpha\tau^2 K + 4\beta\tau^3 K^2 
+ 4\gamma\tau^3 \D K + \cdots \,
\label{Kexpansion}
\ee
where we have used (\ref{Gauss}) to integrate the term $\Om K$. There is
some ambiguity in a K\"ahler potential, but if we insist that $\tK$
and $\K$ both have the asymptotic form $Z\bar{Z} $, and their
difference vanishes asymptotically, then the
ambiguity is resolved. For our selected metrics we can
calculate the quantities $\log\Om$, $K$ and $\D K$ at the origin,
by elementary differentiation, and hence estimate $\tK-\K$ using 
(\ref{Kexpansion}), as a function of $\alpha,\beta,\gamma$. As we explain in the Appendix, we can also 
independently calculate $\tK-\K$ in terms of the on-shell action $S$, 
using the solution of Taubes' equation with the vortex at the origin.
The integral expression for $S$ is given in equation (\ref{Sonshell}).
By comparing these we gain information about the
coefficients $\alpha, \beta, \gamma$, and by varying the parameters of
our metrics, we determine the coefficients precisely. We have found
\be
\label{values_abc}
\alpha = -0.325 \,, \quad \beta = -0.01 \,, \quad \gamma = -0.08 \,.
\ee
The calculations are presented in section 3.

In \cite{ManRic} it was conjectured that $\tO$ is determined from $\Om$
by a Ricci flow, $\Om(\tau)$, starting at $\tau=0$ with $\Om$. 
(The Bradlow parameter $\tau$ is twice the usual `time' encountered 
in Ricci flow.) The Ricci flow equation on a surface, in terms 
of the Gaussian curvature, is
\be
\label{flow_intro}
\frac{\p}{\p \tau} \Om(\tau)= -K(\tau) \Om(\tau) \,.
\ee
One can formally integrate this equation from $\tau=0$, obtaining an 
expansion of the type (\ref{tOexpans}). This approach gives the
correct result for $\tO$ if $\Om$ has constant curvature, but one 
example in \cite{ManRic} showed that the conjecture could not be 
correct in general. 

The conjecture is not entirely wrong. The expansion
(\ref{tOexpans}) is structurally similar to what one gets from Ricci
flow. The leading terms $\Om(1 - \tau K)$ are identical, but the
subsequent coefficients are different. Interestingly, they are
not very different. The comparison is discussed in section \ref{sec_flow}.

The conformal factor (\ref{tOexpans}) rather precisely characterises the
1-vortex moduli space metric $\tilde{g}$ when the background curvature 
is small. Using this we can compare the geodesic motion on the moduli
space to the geodesic motion using $g$. Recall that the latter
represents the motion of a point particle on $\S$, whereas geodesics
using $\tilde{g}$ represent the motion of a vortex on $\S$. A vortex has a
finite size of order $\tau$, so we expect its motion to sample the
background metric over a larger region than a point particle. We find
that, relative to the geodesic of a point particle, the vortex
experiences a force proportional (to leading order in $\tau$) to 
the gradient of the curvature $K$. The additional force is also
proportional to the velocity squared, as one expects for geodesic motion
on the moduli space. As mentioned in section 1, the additional force is 
analogous to the self-force experienced by a body of finite 
size in general relativity. These issues are explored in more detail in 
section \ref{sec_geodesics}. 
  
\section{Calculating the moduli space metric}\news

In this section the Bradlow parameter $\tau$ is set to $1$.
We work with the difference between the moduli space K\"ahler potential 
$\tK$ and the original K\"ahler potential $\K$, both defined on 
$\Sigma$, and using the common coordinate $Z$. As shown in the
Appendix, this has the form
\begin{equation}
\label{eq_for_K}
\tK - \K = S - S_{flat}
\end{equation}
where $S$ is the regularised on-shell action for Taubes' equation, 
which simplifies to
\begin{equation}
S(Z,\bZ)=-\frac{i}{4\pi} \int_\Sigma\,dz\wedge d\bar{z}\, \Omega\,
h \left( 1+ e^h\right) +2a -4\,.
\label{eq:Action}
\end{equation}
Here, $h$ is the solution of the Taubes equation, and $a$ is the constant term in the expansion (\ref{hexpans}) which,
like the function $h$, depends on the vortex location $Z$.
The constant $S_{flat}$ is the on-shell action $S$ evaluated on 
the solution $h_{flat}$ of the Taubes equation on the flat background,
$\Om = 1$. 

From (\ref{Kexpansion}) we expect that $S-S_{flat}$ has the small 
curvature expansion, for $\tau=1$,
\be
S-S_{flat}=2\log\Om + 4\alpha K + 4\beta K^2 
+ 4\gamma \D K + \cdots \,.
\label{Sexpans}
\ee
Our aim is to find the coefficients $\alpha, \beta, \gamma$ and check 
their universality. For this we only need to consider a selection of 
axially symmetric surfaces which are asymptotically planar, and calculate 
$S$ for a vortex at the origin, $Z=0$. To do the calculation we
numerically determine the solution $h$ of Taubes' equation, 
and from its behaviour near the origin extract the coefficient $a$. 
Then we compute the regularised integral $S$ and subtract the constant
$S_{flat}$.

When $\Sigma$ is axially symmetric, the metric (\ref{metric})
becomes 
\be
\label{axial_m}
g = \Om(r)(dr^2 + r^2 d\theta^2)
\ee
where $z=r e^{i\theta}$. For a vortex located at the origin, Taubes'
equation (\ref{Taubes1}) simplifies to the ODE
\be
\frac{1}{\Om}\left(\frac{d^2 h}{dr^2} +\frac{1}{r}\frac{dh}{dr}\right)
+1 - e^h = \frac{4\pi}{\Om} \delta^2(z) \,.
\label{axialODE}
\ee
The expansion of the solution $h(r)$ around $r = 0$ is the simplified 
version of (\ref{hexpans}), 
\begin{equation}
h(r) \sim 2\log\,r + a -\frac{\Omega(0)}{4} r^2 + O(r^4) \,,
\end{equation}
where coefficients of odd powers of $r$ vanish.
The asymptotic form of $h$ for large $r$, which is needed for the numerical
calculations, can be found by noticing that $h$ vanishes as 
$r\rightarrow \infty$, and (\ref{axialODE}) 
reduces to a Bessel equation, with leading asymptotic solution 
\be 
h(r) \sim \frac{\lambda}{\sqrt{r}}\,e^{- \sqrt{\Omega_{as}}\,r } \,,
\ee
where $\Omega_{as}=\lim_{r\rightarrow \infty}\Omega(r)$. Usually $\Omega_{as} = 1$.

Taubes' equation uniquely determines the two asymptotic constants 
$a,\lambda$, but since an explicit solution is
not known, they generally have to be computed numerically.
In the flat case it is possible \cite{deVega} to relate the two 
asymptotic expansions and effectively reduce the problem of solving 
Taubes' equation to a system of transcendental algebraic equations 
relating $a$ and $\lambda$.
A striking, implicitly integrable case is when the metric is related to the 
vortex profile function by $\Omega = e^{-h/2}$. In this case Taubes' equation 
reduces to a $sinh$-Gordon equation \cite{MD}. Also in this case an 
explicit solution is not known, but the two asymptotic expansions of 
$h$ can be related using connection formulae for a particular case of the 
Painlev\'e III ODE (a radial reduction of the $sinh$-Gordon equation) 
and the constant $a$ is uniquely determined, fixing in this way the 
solution.

For the class of metrics of small curvature we are interested in, no 
such methods exist and one has to compute the solution and asymptotic
constants numerically. 

Let us first consider the flat metric with $\Omega=1$. To analyse 
equation (\ref{axialODE}) numerically, we consider a finite interval 
$r\in [\delta, R]$  and then examine the limits $\delta \rightarrow 0$ 
and $R\rightarrow \infty$. Instead of working with the function $h$ we 
remove the logarithmic singularity and consider
\be
u(r)= h(r)-2 \log (r/R)\,.
\ee
The delta function disappears, and $u$ satisfies
\be
\frac{d^2 u}{dr^2} +\frac{1}{r}\frac{du}{dr} + 1 - \frac{r^2}{R^2} e^u=0\,.
\ee
The boundary conditions are $u(\delta) \sim a + 2 \log R 
-\frac{1}{4}\delta^2 + O(\delta^4)$, while for $r = R$ the logarithmic 
term that we added vanishes (but not its derivative) so 
$h(R) = u(R) \sim \frac{\lambda}{\sqrt{R}}e^{-R}$. To obtain the 
solution for $u$ we implemented 
both a shooting and a cooling method, and the two solutions coincide 
within numerical errors. We set $\delta=10^{-4}$ and $R=30$ so $u(R)$ 
effectively vanishes. The values of the constant $a$ and the 
on-shell action $S$ are found to be
\be
a_{flat}=-1.011 \,, \quad
S_{flat}=-1.598 \,.
\ee
These values can also be obtained after some manipulations from 
the asymptotic analysis of \cite{deVega} and they agree within 
numerical errors.

We now consider the family of metrics with conformal factors
\begin{equation}
\Om(r) = \frac{A\, L^4 + B \, L^2 \, r^2 + r^4}{C\,L^4+D\, L^2 \, r^2 +r^4}
\label{metrics}
\end{equation}
with $A,B,C,D$ chosen such that the metric has no zeros or
singularities for $r\in {\mathbb{R}}^+$. These metrics are 
asymptotically planar, with $\Omega_{as} = 1$, while at the origin 
$\Omega(0) = A/C$. 
Thanks to the axial symmetry of $\Om$ the Gaussian curvature can be 
easily obtained using equation (\ref{Gauss}) and takes the form
\be
\label{gaussian_ax}
K = \frac{r \,\Om'(r)^2-\Om(r) \left(\Om'(r)+r \,\Om''(r)\right)}{2\,r\,\Om(r)^3}\,,
\ee
and further differentiation gives $\D K$. Since $\Om$ is quadratic 
around the origin with no linear term, $K(0)= - \Om''(0)/\Om(0)^2$.
In Figure 1 we plot $\Om$ and $K$ as a function of $r$ for the particular 
representative of our family with $A=2, B=3, C=1, D=4$ and $L=1$.

\begin{center}
\includegraphics[scale=0.3]{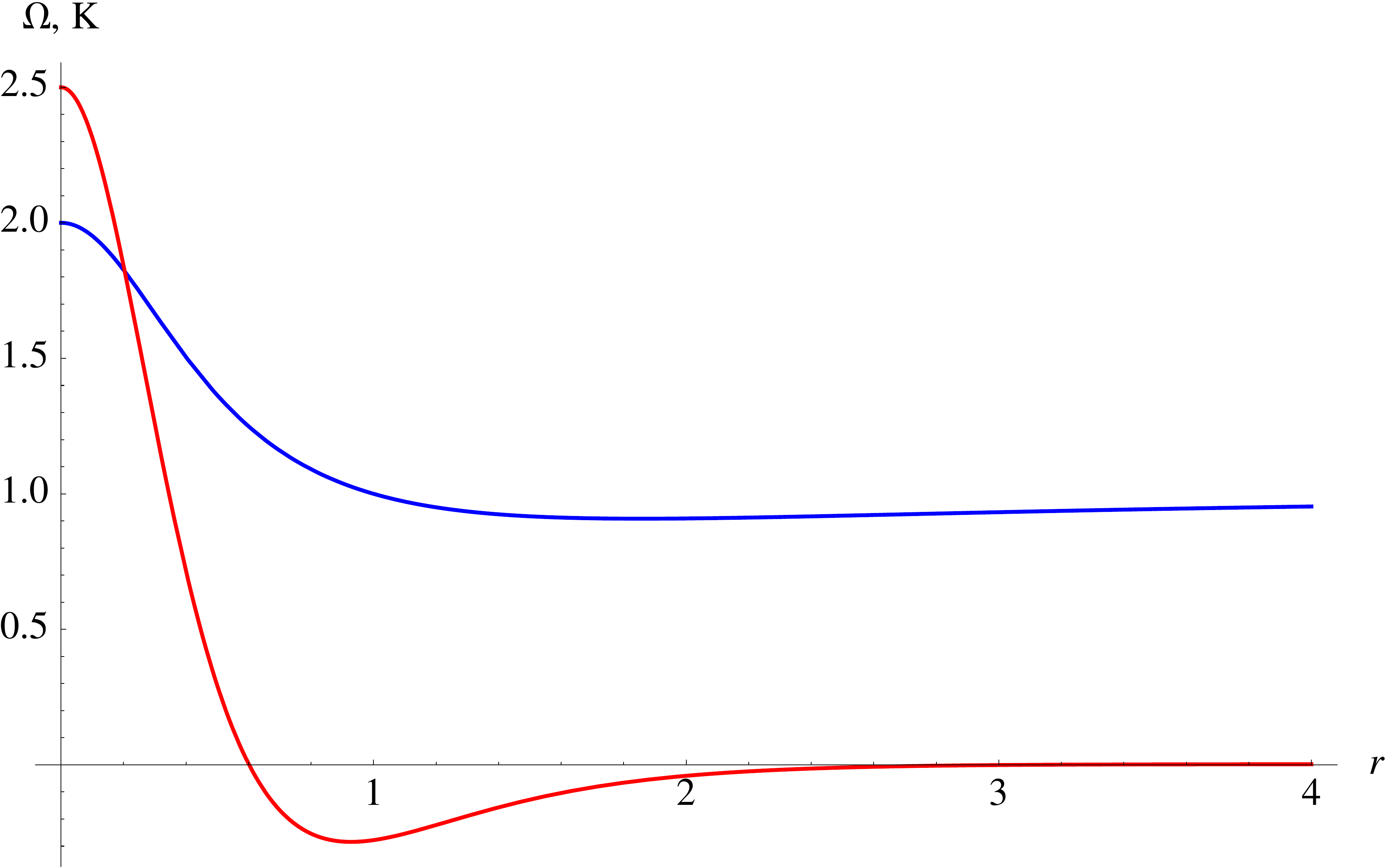}
\begin{center}
{{\bf Figure 1.} {\em The conformal factor $\Omega$ in 
blue and the Gaussian curvature $K$ in red when 
$A=2, B=3, C=1, D=4$ and $L=1$.}}
\end{center}
\end{center}

Recall that the covariant Laplacian $\Delta$ carries a 
factor $1/L^2$, so $K\sim 1/L^2$ while $K^2 , \Delta K \sim 1/L^4$ and 
all higher terms in the expansion (\ref{Sexpans}) come with higher 
powers of $1/L^2$. For the class of conformal factors (\ref{metrics}), and for $Z=0$, equation 
(\ref{Sexpans}) takes the form:
\begin{align}
S-S_{flat} =&\label{ABCD} \,2 \log{\frac{A}{C}}
+ 8 \,\alpha \, \frac{AD-BC}{A^2 L^2}
+ 16\, \beta \, \frac{(AD - BC)^2}{A^4 L^4} \\
&\notag-32\,\gamma\,
\frac{4AC^2 - 4A^2C + 2ABCD + A^2D^2 -3 B^2 C^2}
{A^4 L^4}+O\left(\frac{1}{L^6}\right)\,.
\end{align}
We now fix $L=10$ to suppress all higher order terms. We have also 
increased the value of $L$ to check that the parameters $\alpha,
\beta, \gamma$ do not significantly change.

We may simplify further the class of conformal factors
(\ref{metrics}) by setting $A=C=1$. Now $\Om(0) = 1$, and the curvature
$K$ at the origin is proportional to $B-D$. We consider two 
cases: $B>-2, D=0$ so
\be
S-S_{flat}= -8\,\alpha\, B/L^2
+ 16(\beta  + 6\gamma)B^2/L^4 +O(1/L^6) \,,
\ee 
and $B=0, D>-2 $ so
\be
S-S_{flat}= 8\,\alpha\, D/L^2
+ 16(\beta -2\gamma)D^2/L^4 +O(1/L^6)\,.
\ee
For $D=0$  and $B\in[0,1]$, for each value of $B$ we compute the
solution to Taubes' equation, extract the constant term $a$ near the 
origin and compute the regularised on-shell action $S$. By fitting 
$S-S_{flat}$ with a quadratic polynomial in $B$ we extract the 
values $\alpha=-0.32528$ and $\beta+6\gamma=-0.5125$. Repeating the 
analysis for $B=0$ and $D\in[0,1]$, a quadratic polynomial fit in $D$ 
for $S-S_{flat}$ gives the values $\alpha=-0.32529$ and 
$\beta-2\gamma=0.156$. The value of $\alpha$ in the two cases coincides 
within numerical error and is the first check of the universality of 
the expansion; the other coefficients are determined to be
\be
\beta = -0.011, \quad \gamma = -0.084 \,.
\ee

In Figure 2 we show the numerical data for $S-S_{flat}$ and the
quadratic fits. The two match perfectly for small values of $B,D$ 
but when $B,D\sim 10$ the neglected terms in the expansion (\ref{ABCD}) 
become of order $1$ and our quadratic fits should be replaced by
higher order ones.

\begin{center}
  \begin{tabular}{ l | r }
\includegraphics[scale=0.2]{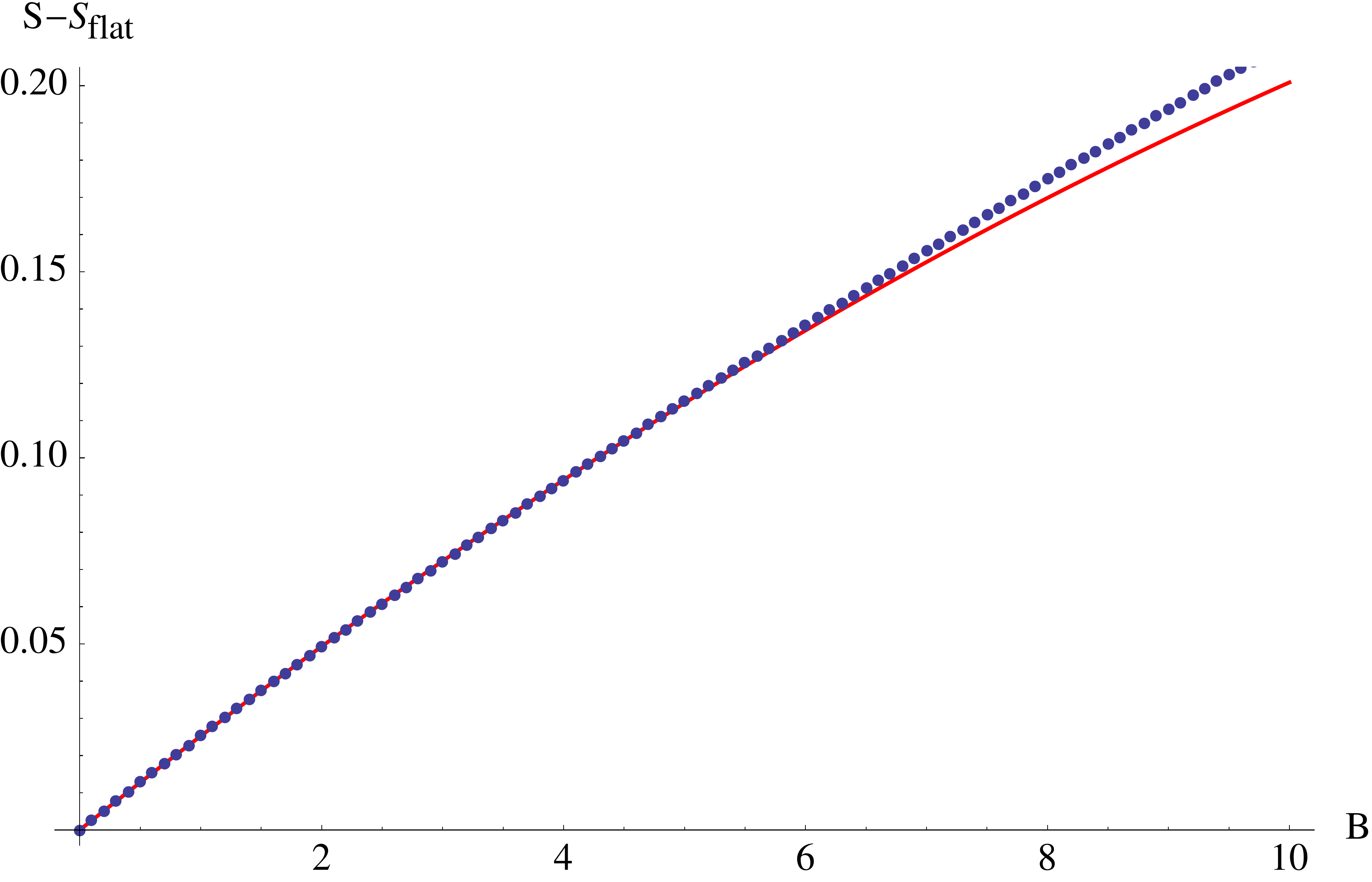} & \includegraphics[scale=0.2]{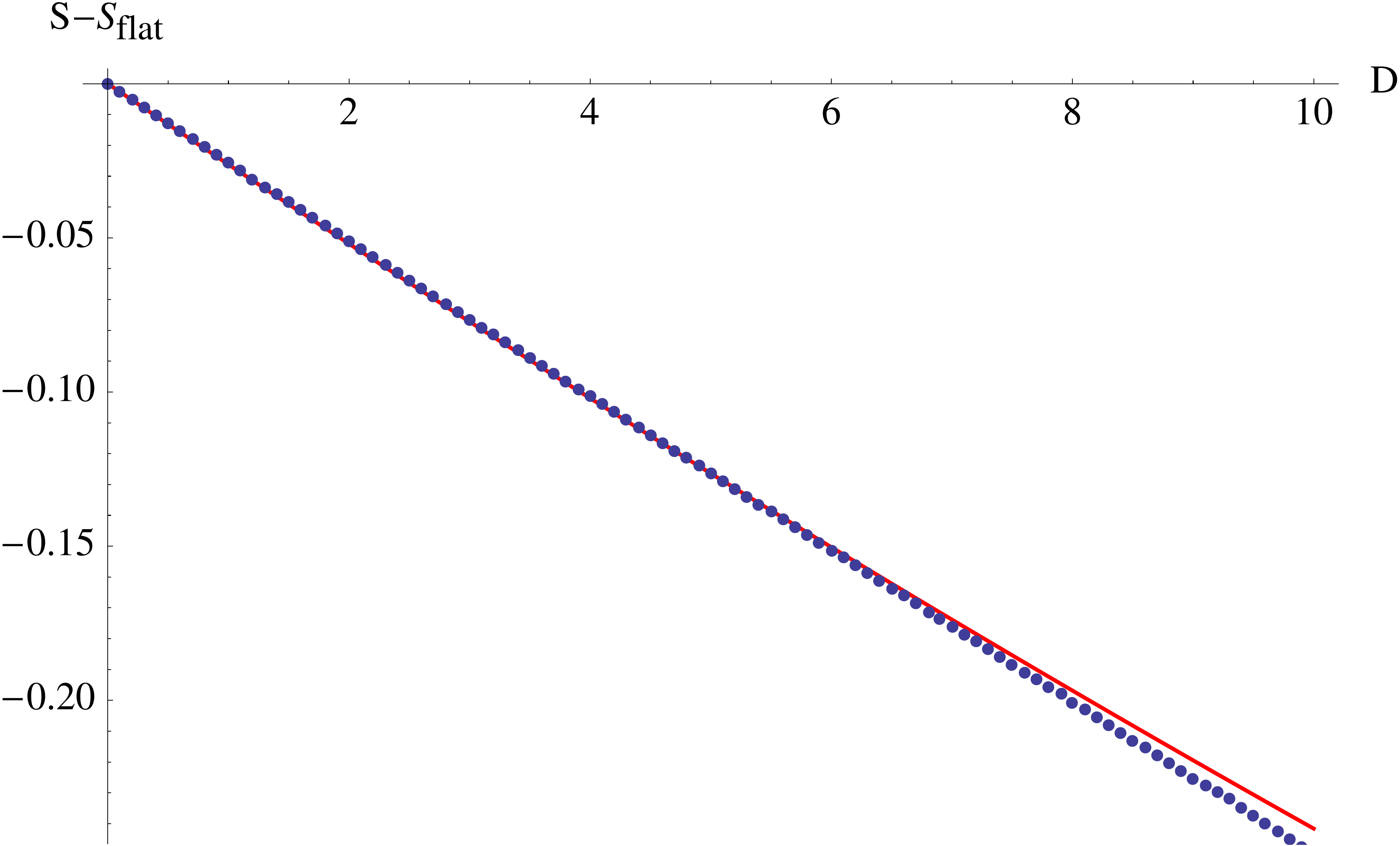}
  \end{tabular}
\begin{center}
  {{\bf Figure 2.} {\em Numerical data in 
blue and quadratic fits in red for $S-S_{flat}$, for the family of metrics with $A=C=1$ and (left) $B\in[0,10], D=0$ and (right) $B=0,D\in[0,10]$.}}
  \label{BDFit}
\end{center}  
\end{center}

To specifically check the universality of the coefficient $\gamma$ 
we reinstate the parameters $A$ and $C$ in (\ref{metrics}) and set 
$B=D=0$. The curvature $K$ now vanishes at the origin but $\Delta K$
does not. We consider the cases: $A>0, C=1$ so
\be
S-S_{flat}= 2 \log A + 128\, \gamma\,(A-1)/(A^3 \,L^4) +O(1/L^6)\,,
\ee 
and $A=1, C>0$ so
\be
S-S_{flat}= -2\log C - 128\,\gamma\,C (C-1)/L^4+O(1/L^6)\,,
\ee
and proceed similarly as before. We first set $C=1$ and vary $A$
in the interval $[1,2]$; for each value of $A$ we numerically solve Taubes' 
equation and calculate the on-shell action $S$. 
Fitting $S-S_{flat}$ with the sum of $2 \log A$ and a cubic 
polynomial in $1/A$ gives a consistent value $\gamma=-0.086$.
Then with $A=1$ and $C\in[1,2]$ we fit with a sum of $-2 \log C$ and 
a quadratic polynomial in $C$ and find $\gamma=-0.085$.
Figure 3 shows the fits in the two cases, compared with data over
larger ranges of $A$ and $C$.
\begin{center}
  \begin{tabular}{ l | r }
  \includegraphics[scale=0.2]{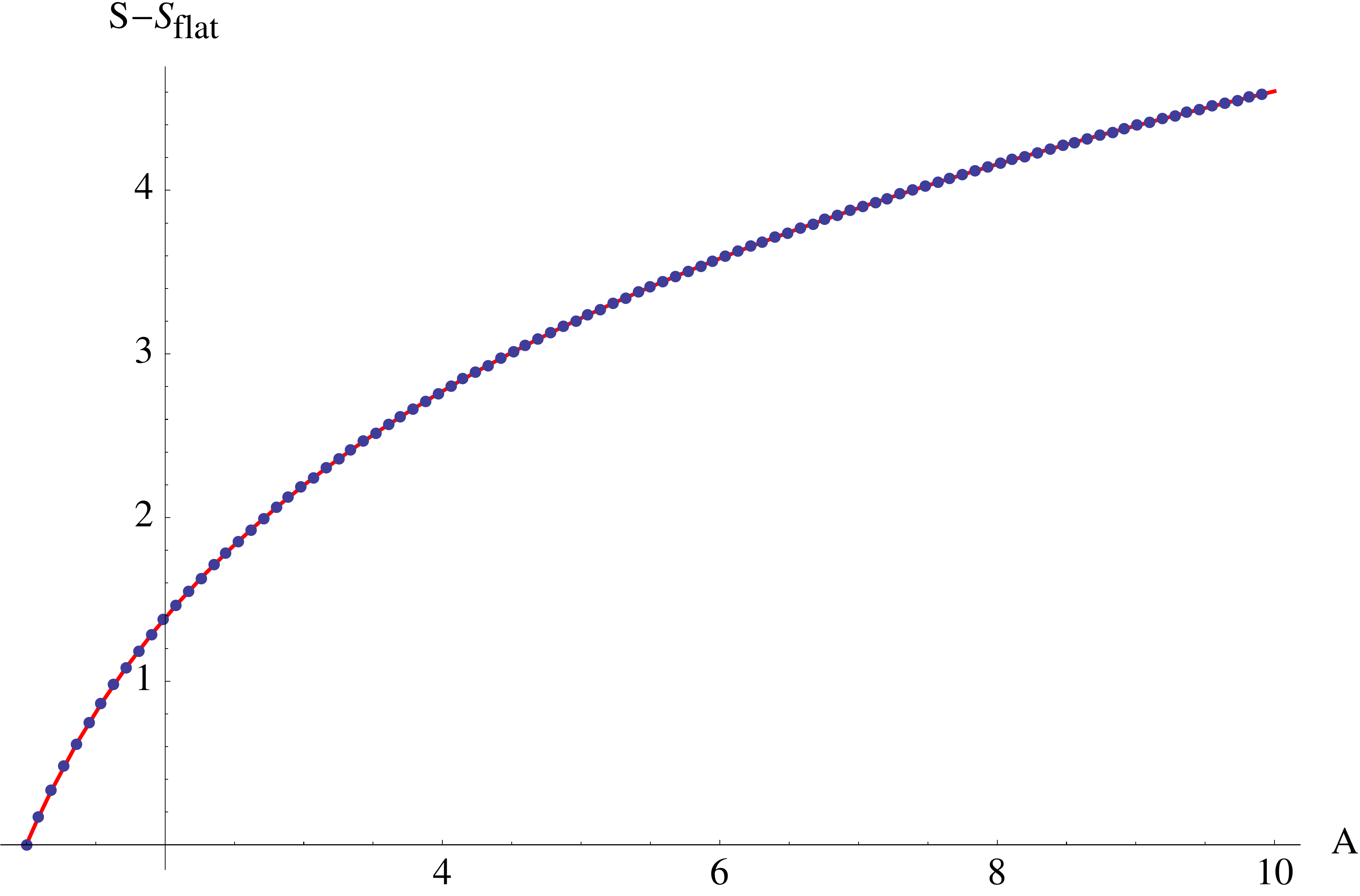} & \includegraphics[scale=0.2]{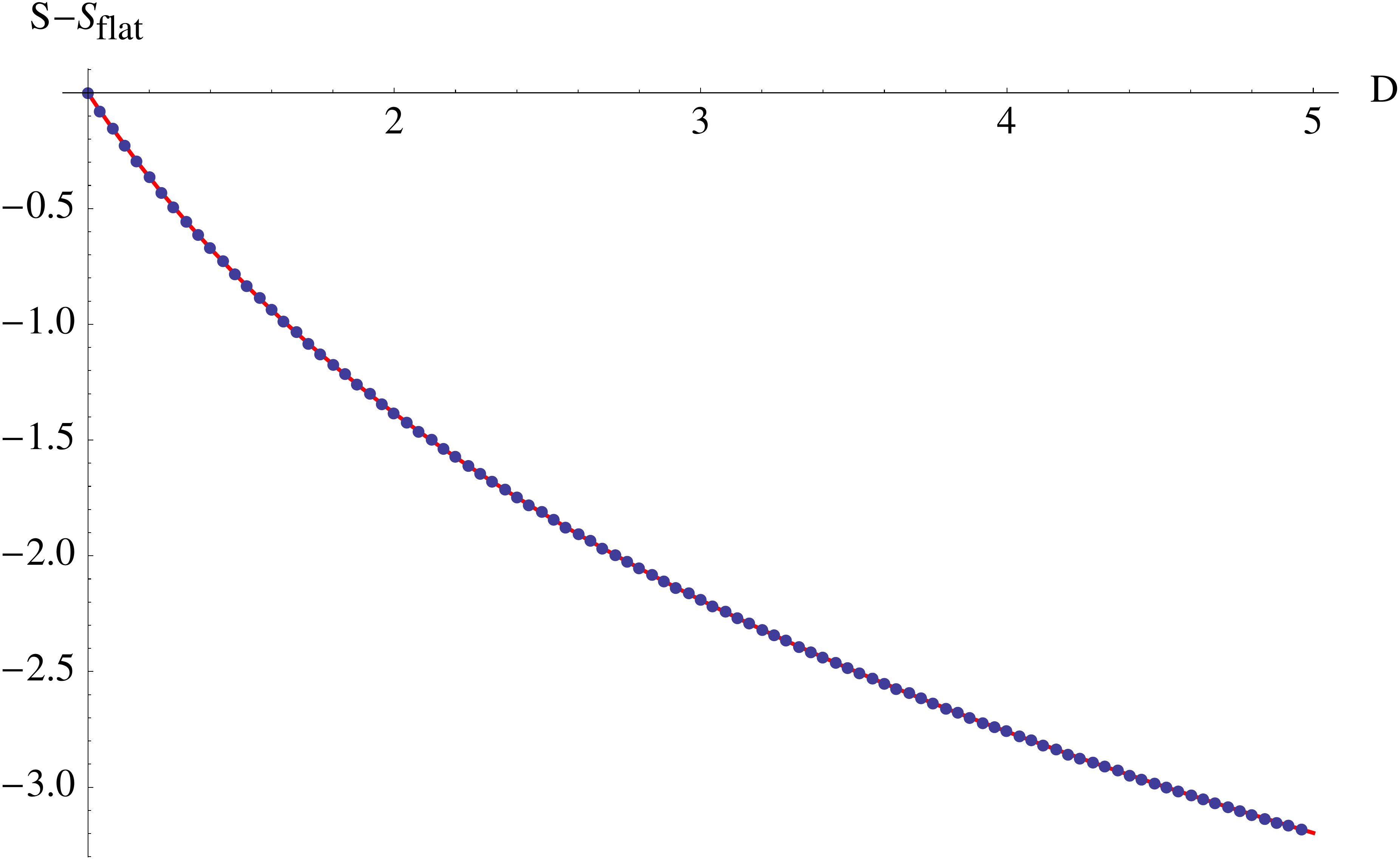}
  \end{tabular}
\begin{center}
 {{\bf Figure 3.} {\em Numerical data in blue and fits in red for 
$S-S_{flat}$, for the family of metrics with $B=D=0$ and (left) 
$A\in[1,10], C=1$ and (right) $A=1,C\in[1,5]$.}}
  \label{ACFit}
\end{center}
\end{center}

To further check the universality of the coefficients within the
family of conformal factors (\ref{metrics}) we also performed a random 
sampling of the multi-dimensional space of parameters 
$A,C\in [0.5,2]^2,\,B,D\in [-1,1]^2$. For $2000$ points in this space 
we computed $h$ and evaluated $S-S_{flat}$. We then performed a 
multi-dimensional nonlinear fit to obtain $\alpha,\beta$ and $\gamma$, 
finding results consistent with the values given above.

Having shown that our coefficients $\alpha,\beta$ and $\gamma$ are 
universal within the chosen family (\ref{metrics}) we have also changed the 
structure of the conformal factors, for example to a family with 
exponential behaviour, sharing the same fundamental features: axial symmetry,
asymptotic planarity, and hierarchical dependence on $L$ for the 
curvature $K$ and its derivatives. We investigated these families as 
above and always obtained similar values for the parameters $\alpha,\beta$ and 
$\gamma$, thus confirming the universality of our expansion.

In summary, our final estimated values of the coefficients, taking 
into account the uncertainty in $\gamma$, are:
\be
\alpha= -0.325\,,\quad \beta = -0.01\,,\quad \gamma=-0.08 \,.
\ee

\section{Comparison with the Ricci flow}\news
\label{sec_flow}

Our underlying idea is that the Taubes equation gives rise to a 
$\tau$-dependent functor mapping a background metric $g$ on $\Sigma$ 
to a moduli space metric $\tilde{g}(\tau)$ on the surface $\M$, which is 
identified with $\Sigma$ as we discussed in section 2. This vortex 
functor has the 
asymptotic expansion (\ref{tOexpans}) in the Bradlow parameter $\tau$ 
(or equivalently in the scaling factor $1/L^2$), and in particular, $\tilde{g}(0)=g$. The terms in this 
expansion include the curvature of the background metric, and other 
higher order, local scalar invariants constructed by acting with powers of 
the covariant Laplacian on powers of the curvature.

The Ricci flow \cite{ricci_flow} on $\Sigma$ also provides a $\tau$-dependent functor 
on the space of metrics. Here, $\tau$ becomes identified with a
multiple of the time-parameter of the flow. This functor is definitely 
local, and in \cite{ManRic} some evidence was given to suggest that 
the vortex functor and the Ricci flow functor are closely related, 
perhaps after reparametrisation of $\tau$. There are clearly
similarities between the two: the Ricci flow expands negatively curved 
regions and shrinks positively curved regions, 
and the same is true for the moduli space metric as a function of $\tau$.
In this section we shall demonstrate that the functors agree to 
lowest order in $\tau$, but differ at higher order in a way which 
is parametrisation independent.

Let $g=g_{ab}(t)$ be a one-parameter family of Riemannian metrics
on a surface. The Ricci flow equation in two dimensions is
\be
\label{rf_old}
\frac{\p}{\p t} g_{ab}=-Rg_{ab} \,,
\ee
where the Ricci scalar $R=2K$ is twice the Gaussian curvature. 
The Ricci flow preserves the conformal class of $g$, so (\ref{rf_old})
can be regarded as a scalar equation for the conformal factor $\Om$.
Moreover, to compare the series solution of 
(\ref{rf_old}) with the expansion (\ref{tOexpans}) of the conformal 
factor $\tO(\tau)$ we shall set $t=\tau/2$, so that (\ref{rf_old}) becomes
\be
\label{rf}
\frac{\p}{\p \tau} \Omega=-K \Omega \,.
\ee
This is the fundamental equation analysed in this section. It implies that
\be
\label{rf2}
\frac{\p}{\p \tau} K=\frac{1}{2}\Delta K+K^2 \,, \qquad 
\left[\frac{\p}{\p \tau}, \Delta \right]=K\Delta \,,
\ee
where $\Delta=g^{ab}\nabla_a\nabla_b = 4\Omega^{-1}\p_z\p_{\bar{z}}$. 
We claim that formally
\be
\label{taylor}
\Omega(\tau)=\Big(1+\tau \gamma_1+\frac{1}{2!}\tau^2\gamma_2+\dots\Big) 
\Omega(0) \,,
\ee
where the $\tau$-independent scalars $\gamma_k$ are homogeneous
polynomials of degree $k$ in $K$ and the Laplacian operator $\Delta$, both 
evaluated at $\tau=0$. Using (\ref{rf2}) we find
\begin{eqnarray}
\label{flow_parameters}
&&\gamma_1=-K \,, \quad \gamma_2=-\frac{1}{2}\Delta K \,,\quad 
\gamma_3=-\Delta\Big(\frac{1}{4}\Delta K+\frac{1}{2}K^2\Big) \,, \\
&& \gamma_4=-\Delta\Big(\frac{1}{8}\Delta^2 K +\frac{1}{4}\Delta K^2
+\frac{3}{4}K\Delta K+K^3\Big) \,,
\quad \cdots\;\;.\nonumber
\end{eqnarray}

Moreover, for all $k \ge 2$, $\gamma_k$ is of the form of $\Delta$
acting on a homogeneous polynomial in $(K, \Delta)$ of degree $k-1$. 
This can be seen by an inductive argument: if
\be
\frac{\p^k}{\p \tau^k}\Omega=-\Delta(\rho_k)\Omega 
\ee
is the $(k-1)$st derivative of (\ref{rf}), then (\ref{rf}) and 
(\ref{rf2}) imply that
\begin{eqnarray}
\frac{\p^{(k+1)}}{\p \tau^{(k+1)}}\Omega&=&\Big(-K\Delta(\rho_k)-
\Delta\Big(\frac{\partial{\rho_k}}{\partial \tau}\Big)
+ \Delta(\rho_k)K\Big)\Omega
\nonumber \\
&=&-\Delta \Big(\frac{\partial{\rho_k}}{\partial \tau}\Big)\Omega \,.
\end{eqnarray}
The coefficients (\ref{flow_parameters}) are most easily obtained by
using this formula, and setting $\tau=0$. 

The expansion (\ref{taylor}) can also be derived using  
Picard iterations of conformal rescalings. Recall that the Gaussian 
curvature of a conformally rescaled metric $\hat{g}=\omega g$ is
\be
\label{conf}
\hat{K}=\omega^{-1}\Big(K-\frac{1}{2}\Delta(\log \omega)\Big) \,.
\ee
Let $\Omega_0=\Omega$ be a conformal factor  on $\Sigma$ which does
not depend on $\tau$. Define a sequence $\Omega_0, \Omega_1, \dots$ 
of $\tau$-dependent conformal factors by
\be
\label{piccard}
\Omega_{n+1}(\tau)=\Omega_0-\int_0^{\tau} K_n(s)\;\Omega_n(s)ds \,,
\ee
where $K_n$ is the Gaussian curvature of $\Omega_n$. The limit of this 
sequence is the solution to the Ricci flow equation (\ref{rf}), by 
Picard's theorem. At this stage we do not assume that $\tau$ is
small. At each step of the iteration we find $\Omega_n(\tau)
=\omega_n(\tau)\Omega_0$, where $\omega_n$ is some conformal 
factor of the form
\be
\omega_n=1+\tau\gamma_1+\dots+\frac{\tau^n}{n!}\gamma_n \,.
\ee
We can therefore find $K_n$ using formula (\ref{conf}).
The Picard expansion yields
\be
\omega_{n+1}(\tau)=1-\tau K+\frac{1}{2}\Delta{\int_0}^\tau 
\log{(\omega_n(s))} \, ds \,,
\ee
where $\Delta$ is the Laplacian associated with $\Omega_0$. If we now 
assume that 
$\tau$ is small, and expand (in $s$ which is also small) the logarithms 
in the integrand above keeping all terms up to $s^n$, then the successive
iterations agree with (\ref{taylor}), in the sense that the $n$th iteration
reproduces the first $n+1$ terms in (\ref{taylor}). Moreover $\omega_{n+1}
(\tau)$ differs from  $\omega_n(\tau)$ by a monomial of the form
$\tau^{n+1}\gamma_{n+1}/n!$, and so the $n$th Picard iteration preserves
the first $n$ terms in the expansion.

To compare the Ricci flow expansion (\ref{taylor}) with the 
expansion (\ref{tOexpans}) of the conformal factor $\tO(\tau)$ on the moduli 
space, write out the first four terms in (\ref{taylor}) as 
\be
\label{conf_ricci}
\Omega(\tau)=\Big(1-\tau K-\frac{1}{4}{\tau^2}\D K
-\frac{1}{24}\tau^3(\D^2K+2\D K^2)+\dots\Big)\Omega(0) \,,
\ee
and set $\Om(0)=\Om$.
This expansion is of the form of (\ref{tOexpans}) with 
\be
\alpha=-\frac{1}{4} \,, \quad \beta=-\frac{1}{12} \,,\quad
\gamma=-\frac{1}{24} \,. 
\ee
These coefficients differ from those in (\ref{values_abc}), obtained 
by calculating the K\"ahler potential on the moduli
space. Therefore the Ricci flow does not integrate
to the functor giving the moduli space conformal factor $\tO(\tau)$, 
although it does not differ from it greatly for small values of $\tau$.

One may ask if $\tO(\tau)$ can be regarded as the 
solution of a different flow equation in $\tau$, which we refer to as vortex 
flow in contrast to Ricci flow. Let us assume that this is a local 
geometric flow on $\Sigma$. It has to be of first order in 
$\p / \p \tau$ because the background metric $g$ on $\Sigma$ specifies 
the moduli space metric uniquely, so there is no room for more 
initial data. As the first term in the expansion (\ref{tOexpans}) agrees
with that of the Ricci flow, the RHS of (\ref{rf}) has to be modified in a
non-autonomous way. To the next order in $\tau$, the vortex flow is
\be
\frac{\p}{\p \tau}\tO=(-K + c\tau \Delta K)\tO \,,
\label{vortflow2}
\ee
for some constant $c$. Using the Picard method to iterate the 
resulting integral equation, we find this vortex flow reproduces 
(\ref{tOexpans}) up to quadratic terms in $\tau$, provided one chooses
$c=2\alpha + \half$. The cubic terms now need to be corrected by a 
further modification of the RHS of (\ref{vortflow2}) and so forth. The
problem essentially reduces to finding an operator $f(\tau\Delta)$ 
with $f(0)=-Id$ such that the vortex flow  is
\be
\frac{\p}{\p \tau}\tO=[f(\tau\Delta)(K)]\tO \,.
\ee
The iterative analytical procedure for constructing the operator $f$ 
should involve a Green's function for the linearised Taubes equation, 
but this function is not known in closed form.

On the basis of the dimensional analysis carried out in section 2, 
the vortex flow expansion (\ref{tOexpans}) could in principle contain 
terms involving the norm of the gradient of the Gaussian curvature. 
These terms can not arise at orders lower then four, but already the 
coefficient of $\tau^4$ might involve a constant multiple of 
$\Delta|\nabla K|^2$. On the other hand we have established that the 
Ricci flow expansion contains no such terms, and it may be possible to 
prove their absence in the vortex flow.  

\section{Particle geodesics and vortex paths}\news
\label{sec_geodesics}

On a simply-connected surface of constant curvature $K$ the 1-vortex 
moduli space metric is known to be $\tilde{g} = (1 - \tau K)g$, just 
a constant multiple of $g$. It can be constructed exactly without using 
the expansion (\ref{tOexpans}), but instead relying on symmetry arguments 
\cite{manton_stat, Sam}. Thus a vortex moves along a
geodesic of the original surface, and its path coincides with that of a point 
particle, although a vortex and a point particle with the same
initial position and kinetic energy will usually reach the same destination at
different times. The vortex and particle have different inertial 
masses, so their velocities differ even if they have the same kinetic 
energy. On a sphere the vortex has smaller mass -- measured by 
the conformal factor -- than the particle. On a hyperbolic plane the 
vortex has larger mass.  

This picture changes if the curvature of the background metric is not 
constant, as a vortex of finite size is then affected by the
background metric in a larger region than a point particle. Thus we 
expect the vortex and particle paths with the same initial conditions to be
different. The acceleration of the vortex will differ from that of the
particle -- an effect which can be attributed to an additional force. In this section we shall investigate 
this effect.

To first order in the Bradlow parameter $\tau$, the moduli space metric is
\be
\label{manton_zero}
\tilde{g}(\tau)=\Big(1-{\tau K}+O(\tau^2)\Big)g \,,
\ee
where $K$ is the  Gaussian curvature of the background metric $g$.
The point particle follows the geodesics of $g$, and in the geodesic 
approximation the vortex moves along the geodesics of $\tilde{g}$. 
If $x^a=(x, y)$ are local real coordinates on $\Sigma$, then the 
affinely parametrised geodesics $x^a(s)$ of $\tilde{g}$ are integral 
curves of the system of ODEs
\be
\label{new_geodesics}
\ddot{x}^a+\Gamma_{bc}^a \dot{x}^b\dot{x}^c
=-\dot{x}^b\Upsilon_b \dot{x}^a+\frac{1}{2}
g_{bc}\dot{x}^b\dot{x}^c \Upsilon^a, \quad a=1, 2
\ee
where overdots denote derivatives w.r.t. $s$ and $\Gamma^a_{bc}$ is the 
Levi--Civita connection of the metric $g$. The quantity
$\Upsilon_a$ on the RHS is
\be
\Upsilon_a=\nabla_a\log{(1-\tau K +O(\tau^2))} 
\cong -\tau\nabla_a K+O(\tau^2) \,,
\ee
which vanishes when $\tau=0$ or if $K$ is constant.
Using the K\"ahler coordinates on $\Sigma$ such that the metric $g$ is 
given by (\ref{metric}), the vortex geodesics are approximated by 
integral curves of the equation
\be
\ddot{z}+(\Omega^{-1}{\p_z \Omega})\dot{z}^2=\tau (\p_z K)\dot{z}^2
+O(\tau^2)
\label{zgeod}
\ee
and the complex conjugate of this. The equation of motion for a
vortex differs from the equation for a point particle
by the terms on the RHS of (\ref{new_geodesics}) or (\ref{zgeod}). 
Therefore, to first order in $\tau$, the vortex experiences a force 
proportional to the gradient of the curvature, and also proportional 
to the velocity squared. 

To compare vortex paths with the paths of point particles 
we only require unparametrised geodesics. 
Eliminating the affine parameter $s$ between the two equations 
(\ref{new_geodesics}) leads, for each $\tau$, to a single second 
order ODE for $y$ as a function of $x$,
\be
\label{unparametrised}
\frac{d^2 y}{d x^2}=\Gamma^1_{22}(\tau) \Big(\frac{d y}{d x}\Big)^3
+ (2\Gamma^1_{12}(\tau)-\Gamma^2_{22}(\tau)) \Big(\frac{d y}{d x}\Big)^2
+(\Gamma^1_{11}(\tau)-2\Gamma^2_{12}(\tau)) \Big(\frac{d y}{d x}\Big)
-\Gamma^2_{11}(\tau) \,,
\ee
where $\Gamma^a_{bc}(\tau)
=\Gamma^a_{bc}-\tau(\delta^a_b\nabla_c K+\delta^a_c\nabla_b K - 
 g_{bc}\nabla^a K)/2 +O(\tau^2)$.

We shall solve this equation numerically, for $(\Sigma, g)$ a
surface of revolution with metric (\ref{axial_m})
and Gaussian curvature $K$ given by (\ref{gaussian_ax}).
The ODE (\ref{unparametrised}) becomes, to first order in $\tau$,
\be
\label{radial_ode}
\frac{d^2 r}{d \theta^2}=r+\frac{r^2}{2}\frac{\p_r\Omega}{\Omega}
-\frac{\tau}{2}r^2 \p_r K
+\Big( \frac{2}{r}+\frac{1}{2}\frac{\p_r \Omega}{\Omega}
-\frac{\tau}{2}\p_r K\Big)\Big(\frac{d r}{d \theta}\Big)^2 \,.
\ee

As an example we consider a conformal factor of the form
(\ref{metrics}), with $L=1$, and choose the constants $A, B, C, D$ so that
\be
\Omega=\frac{2+7r^2+r^4}{1+r^2+r^4} \,.
\label{ExMetric}
\ee
This metric is asymptotically planar, and the radial plots of the 
Gaussian curvatures of the background metric $g$ and 
the modified metric $\tilde{g} = (1-K)g$ as functions 
of $r$ are given in Figure 4. Note that both curvatures integrate to zero as anticipated in equation (\ref{GauBonNC}).
\begin{center}
\includegraphics[width=10cm,height=6cm,angle=0]{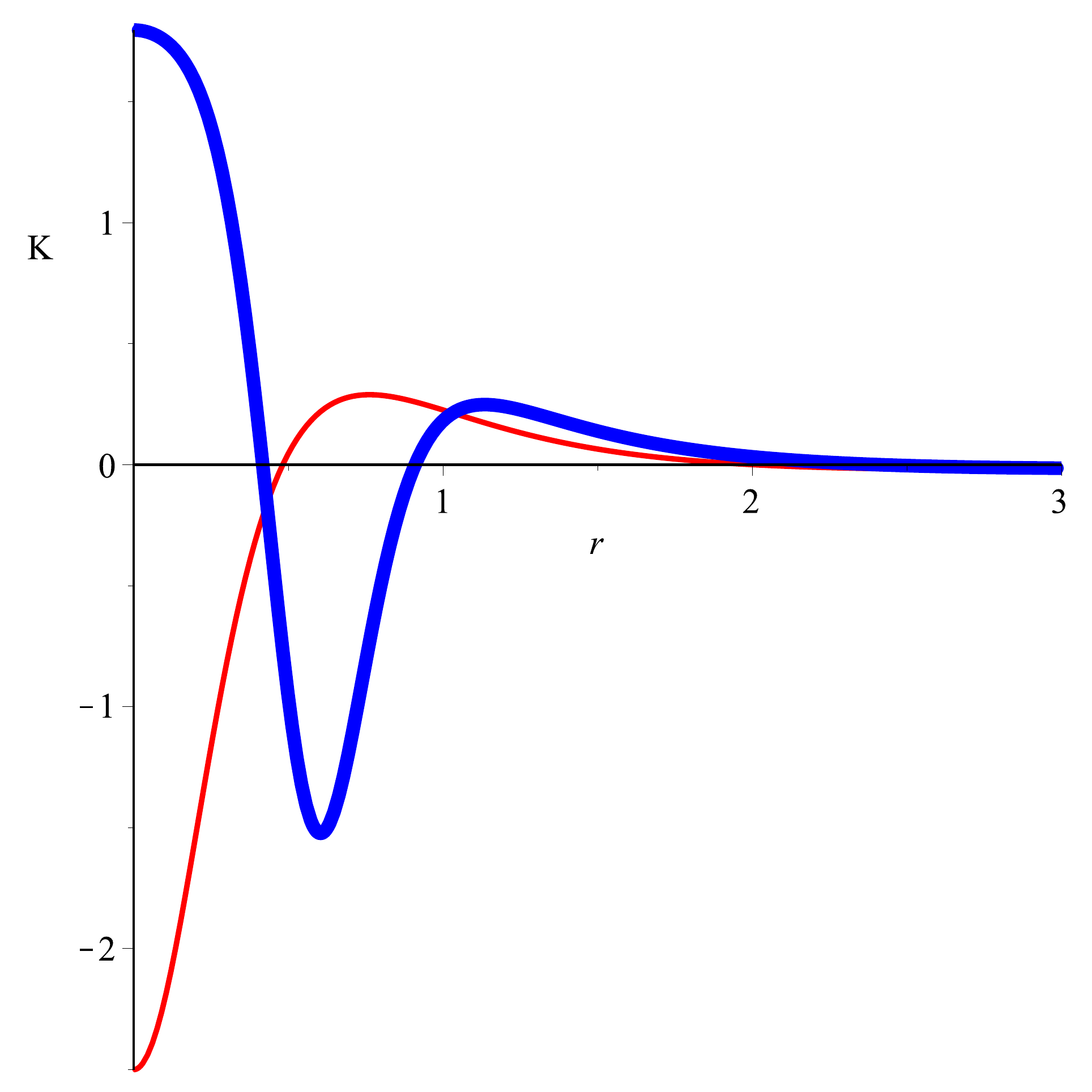}
\begin{center}
{ {\bf Figure 4.}  {\em Gaussian curvatures of the asymptotically flat 
surface of revolution with conformal factor (\ref{ExMetric})
(thin red curve) and the modified vortex metric (thicker blue curve).}}
\end{center}
\end{center}

We have set $\tau=1$ here, but still
work to first order in $\tau$. The resulting curvature is 
not small in the sense explained in section 2, and thus the deviation 
effects are approximate and exaggerated on the Figures below. 
However, the nature of these effects would not change
for small $\tau$ or equivalently small curvature, but the differences 
in the Figures would be less visible. 

For the chosen conformal factor the radial force $\p_r K$ in the
ODE (\ref{radial_ode})  is
attractive for $r\in (0, r_1)$ and $r\in(r_2, \infty)$ and
repulsive for $r\in (r_1, r_2)$, where $r_1\approx 0.76, r_2\approx 2.89$
(Figure 5).
\begin{center}
\includegraphics[width=10cm,height=6cm,angle=0]{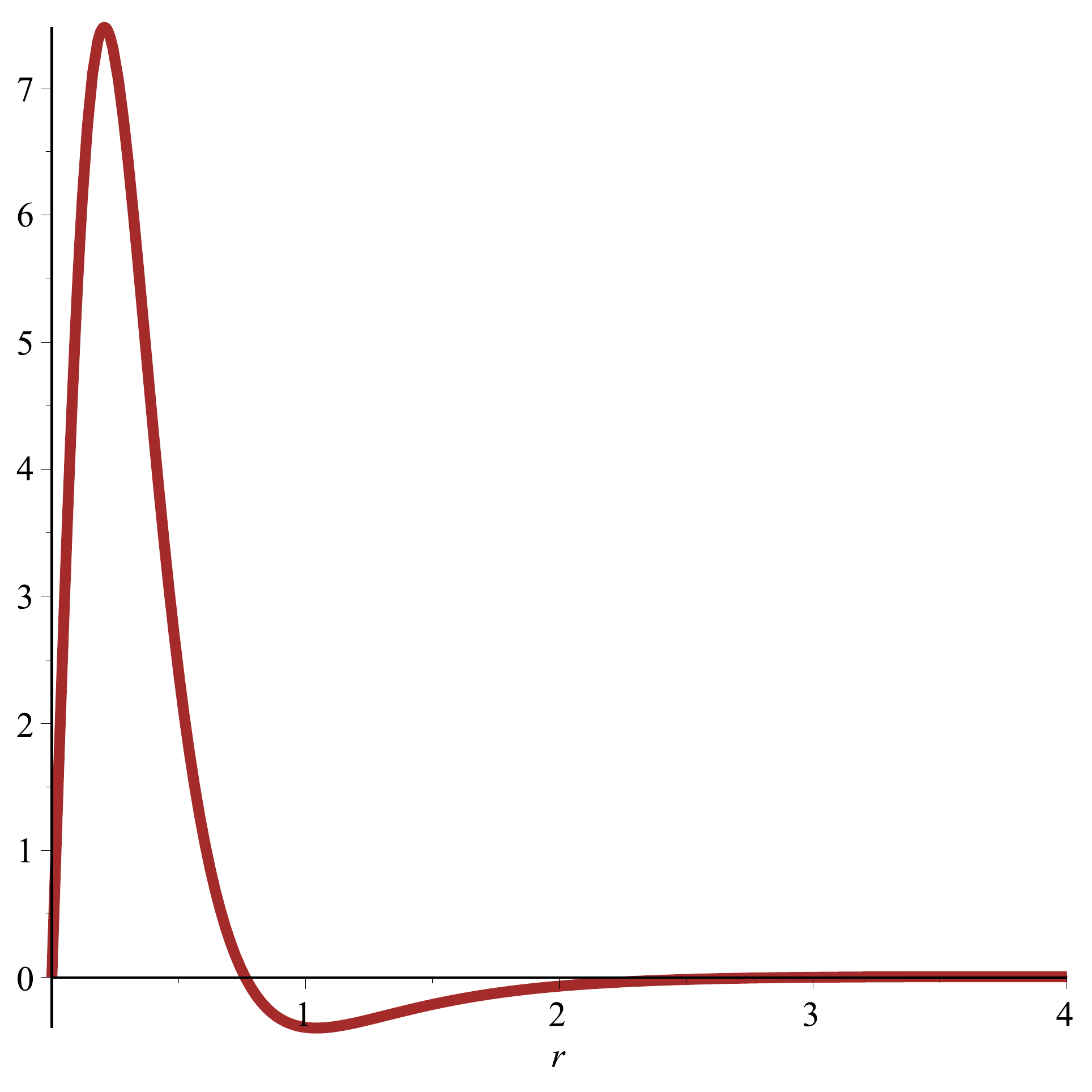}
\begin{center}
{ {\bf Figure 5.}  {\em The radial force $\p_r K$ as a function of $r$.}}
\end{center}
\end{center}
Figure 6 shows both the point particle geodesic 
($\tau=0$) and the vortex geodesic ($\tau=1$) approaching from the same point in the 
asymptotically flat region and in the same initial direction.
The paths coincide initially, but the vortex path moves apart from the
particle path in the region where the curvature gradient 
is large.  The paths then approach straight lines again in the
asymptotically flat region, but in different directions.
The vortex path can either extend further out in the radial direction, or
bend towards the origin.  This depends on the sign of the radial force
$\partial_r K$ in the region where its absolute value becomes large.
This in turn depends on the initial conditions.
\begin{center}
\begin{tabular}{ l | r }
\includegraphics[width=6cm,height=6cm,angle=0]{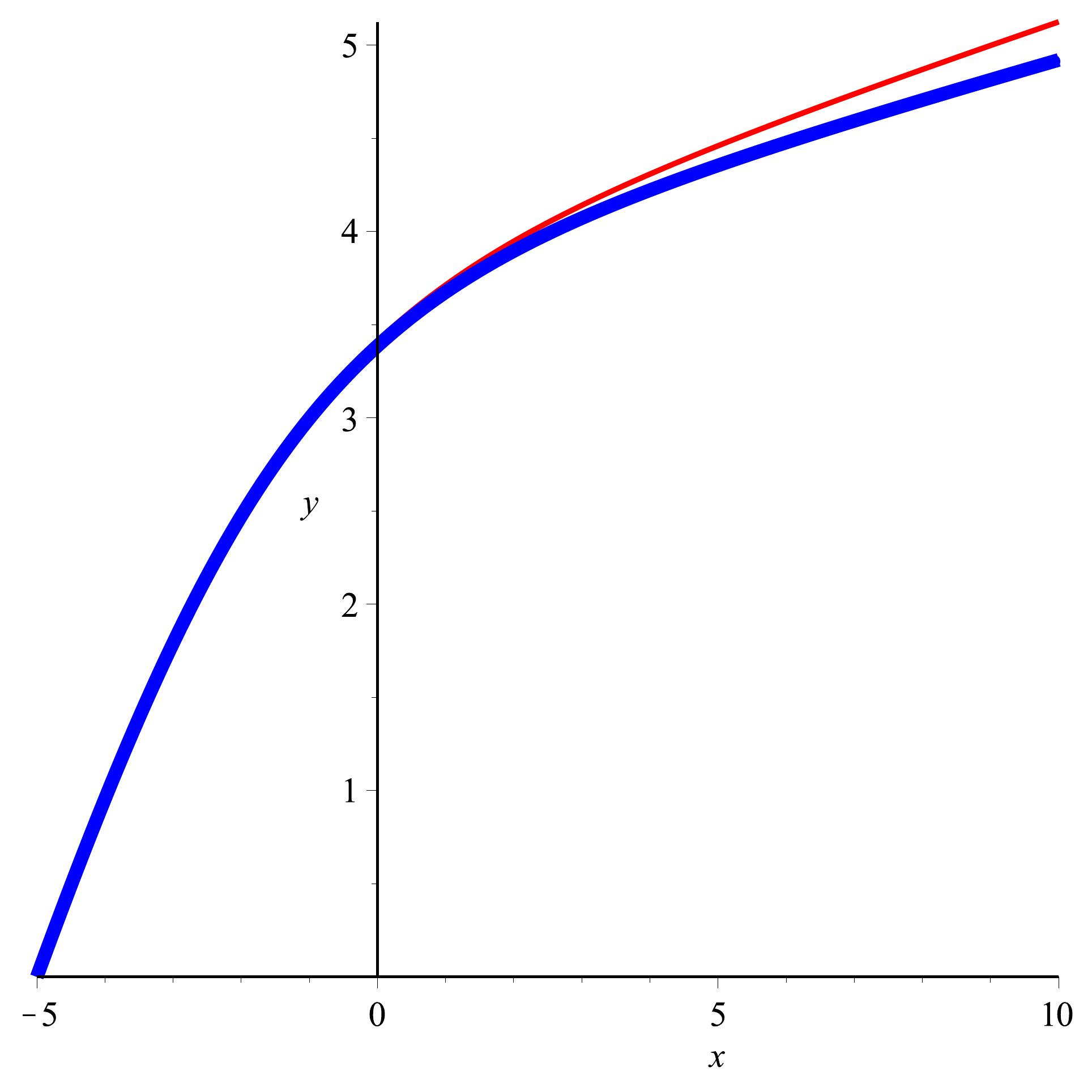}&
\includegraphics[width=6cm,height=6cm,angle=0]{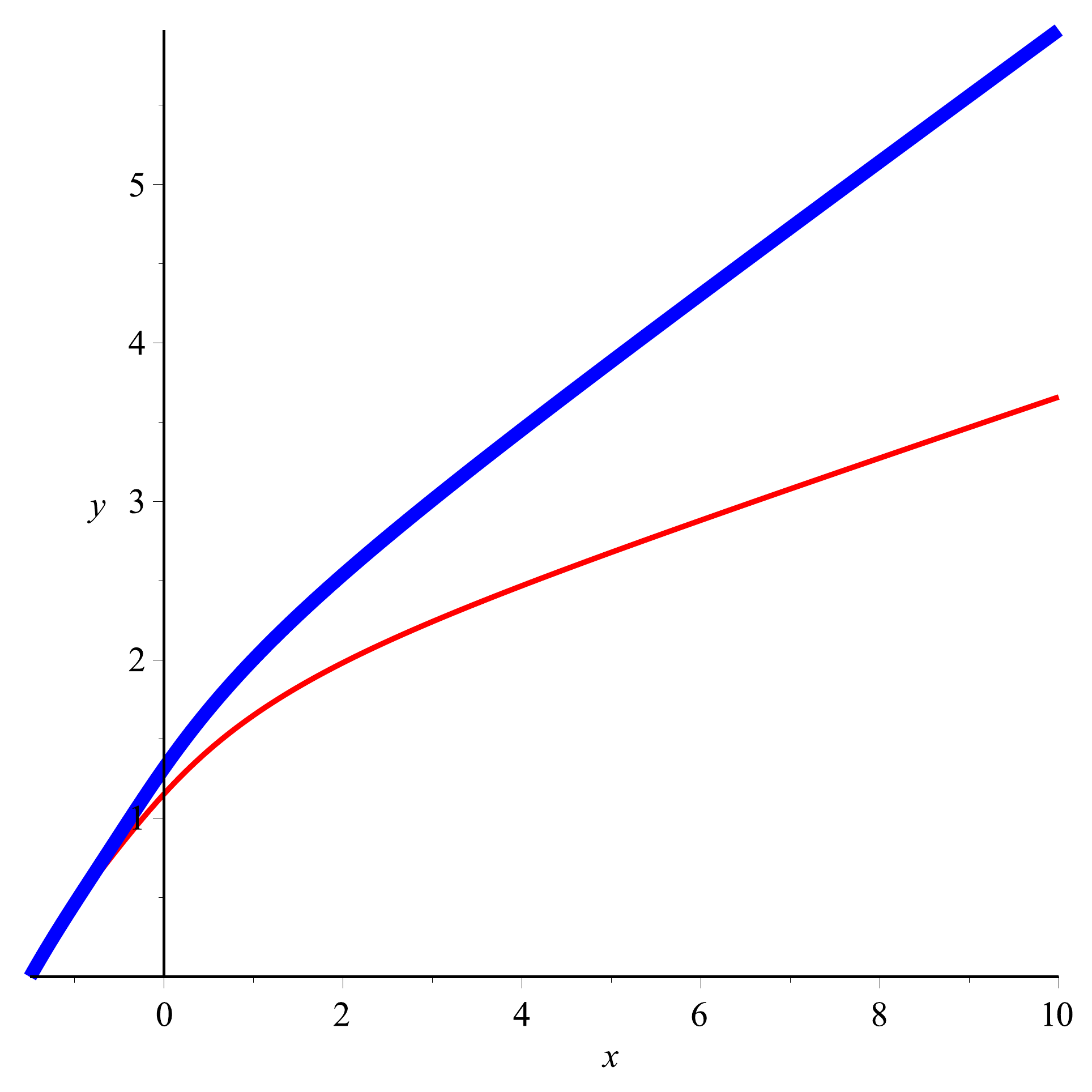}
\end{tabular}
\begin{center}
{{\bf Figure 6.}
{\em Particle geodesics (thin, red), and 
vortex geodesics (thicker, blue) with the same initial conditions:
$\{y(-5)=0, \, y'(-5)=1\}$ (left) and $\{y(-1.5)=0, \, y'(-1.5)=1\}$
(right).
}}
\end{center}
\end{center}

Figure 7 shows  the geodesics of a point particle and a vortex going 
through the same initial and final points. These points are in 
the region of small negative curvature of both the background metric $g$
and the approximate moduli space metric $\tilde{g}$. The vortex path extends 
further out in $r$ than the point particle path. 
\begin{center}
\includegraphics[width=6cm,height=6cm,angle=0]{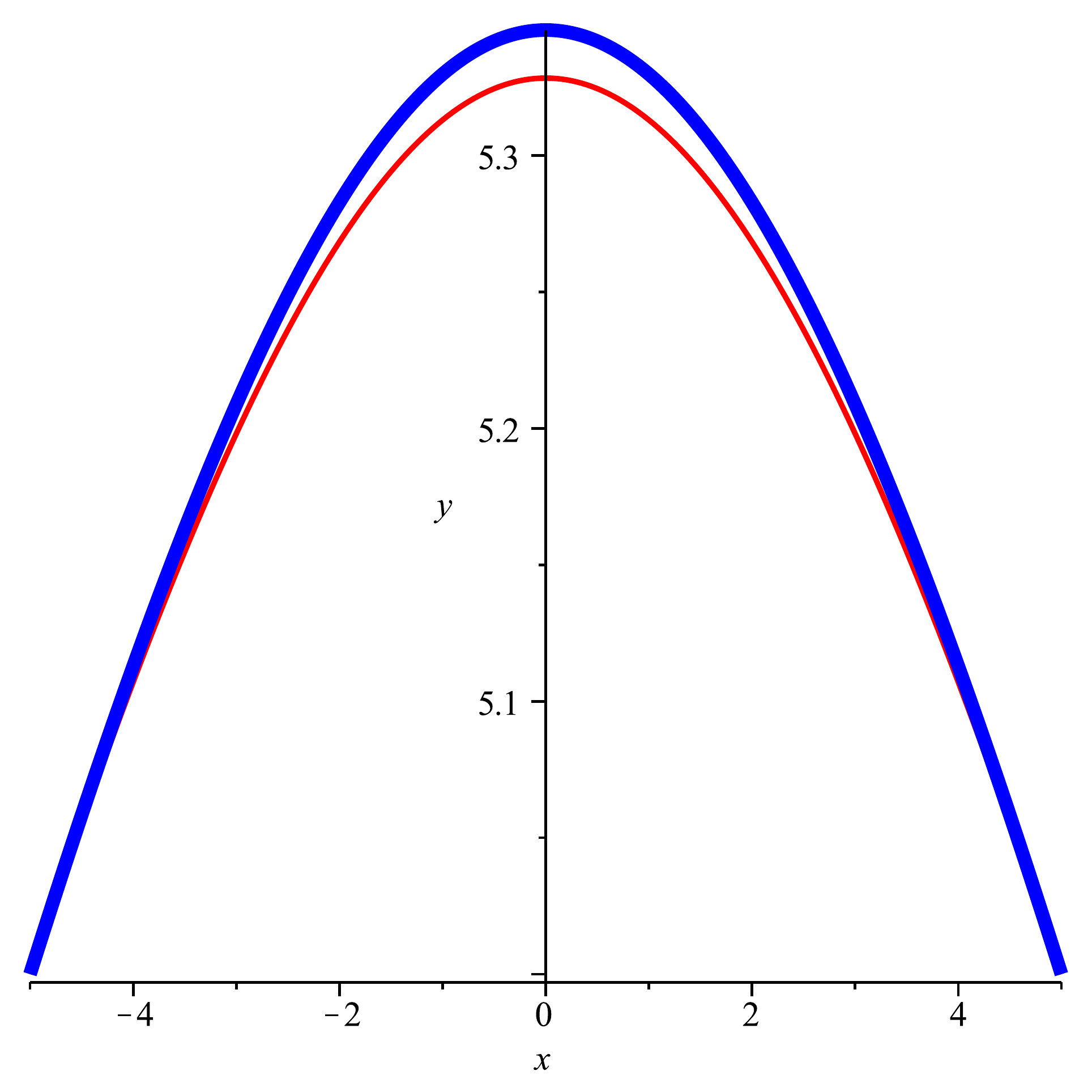}
\begin{center}
{ {\bf Figure 7.}  {\em Geodesic of a point particle (thin, red), and 
a vortex (thicker, blue) passing through the same initial and final points 
$(-5, 5)$ and $(5, 5)$.}}\end{center}
\end{center}

Let us finally consider an example of vortex motion on a compact surface.
We assume that our analysis leading to the numerical values
(\ref{values_abc}) of the coefficients in the expansion of the
conformal factor of the moduli space, (\ref{tOexpans}), 
applies in the compact case, although we have not established this. The 
first term in the expansion is certainly universal. To see the
curvature effects, we choose to work with the ellipsoid of revolution 
\be
x^2+y^2+ \frac{\zeta^2}{b^2}=1 \,,
\ee
whose Riemannian metric induced from ${\mathbb{R}}^3$ by eliminating
$\zeta$ and setting $x+iy=re^{i\theta}$ is
\be
g=\frac{1-(1-b^2)r^2}{1-r^2}dr^2+r^2d\theta^2 \,.
\ee
As before, the vortex path deviates from a point particle path in 
the region of large curvature gradient, which for an ellipsoid with $b<1$ is
close to the equator. This is illustrated in Figure 8.
\begin{center}
\includegraphics[width=6cm,height=6cm,angle=0]{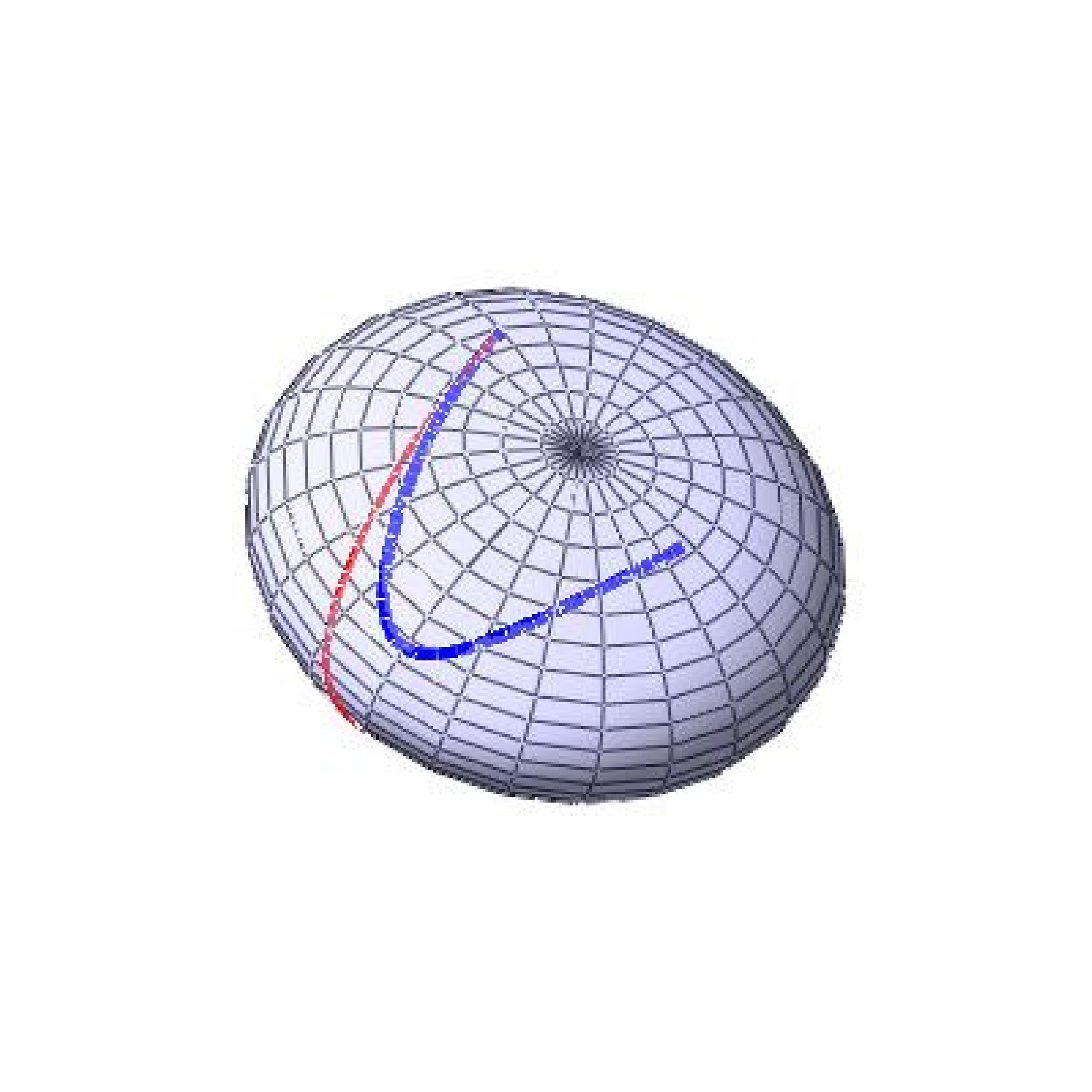}
\begin{center}
{ {\bf Figure 8.}  {\em Point particle geodesic (thin, red) and vortex geodesic 
(thicker, blue) on the ellipsoid of revolution $x^2+y^2+4\zeta^2=1$.}}
\end{center}
\end{center}

\section{Conclusions}
We have studied the moduli space metric for one Abelian Higgs vortex on a surface of small curvature, or equivalently, for a vortex of small size on a fixed surface. This work complements a previous study of the moduli space metric close to the Bradlow limit where the vortex is of comparable size to the whole surface, and is close to dissolving \cite{Manton:2010sa}.
The moduli space metric has a universal expansion in the Bradlow parameter (controlling the size of the vortex) involving the background metric, its curvature and the Laplacian operator acting on powers of the curvature. We are not able to calculate the coefficients in this expansion analytically, but have calculated the first few of them numerically using a family of asymptotically planar, circularly symmetric surfaces, and have checked their universality.

The moduli space metric as a function of the Bradlow parameter has some similarity to the solution of the Ricci flow as a function of time with the underlying metric as initial data. Their Taylor expansions agree to first order but not to higher order. We have found a modification of the Ricci flow which reproduces correctly the expansion in the Bradlow parameter to second order and further improvements are possible.

We have investigated vortex motion by calculating geodesics on the moduli space and have compared these to geodesics on the original surface, which model the motion of point particles. They differ because of the finite vortex size. The leading effect is an additional force acting on a vortex, proportional to the gradient of the curvature of the surface. This additional force has some analogy with the self-force experienced by a finite-size body moving in a gravitational background.

\section*{Acknowledgements}
D. D. is grateful for the support of European Research Council Advanced Grant No. 247252, Properties and Applications of the Gauge/Gravity Correspondence.
M. D. thanks Ian Anderson for his helpful assistance with the MAPLE 
DifferentialGeometry package. 

\section*{Appendix. The regularised Taubes action }
\setcounter{equation}{0}
\appendix
\def\theequation{\thesection{A}\arabic{equation}}

\label{appendix}
Here we construct an integral formula for the K\"ahler potential 
on the 1-vortex moduli space, for a vortex moving on the Riemann surface 
$\Sigma$ with conformal factor $\Om(z,\bar{z})$. We assume that 
$\Sigma$ is asymptotically planar with $\Om = 1$ at infinity.

We first note \cite{CM} that Taubes' equation for a vortex located at
$Z$ arises as the variational equation for the action
\begin{equation}
S= \lim_{\epsilon \to 0} 
\left\{ \frac{i}{2\pi}\, \int_{\Sigma_\epsilon} dz \wedge d\bar{z}
\left[ 2 \partial_z h\,\partial_{\bar{z}}h+\frac{\Omega}{\tau} 
\left(e^h-1-h\right)\right] +4 a +4 \log \epsilon \right\} \,,
\label{eq:TaubesAction}
\end{equation}
where $\Sigma_\epsilon =\Sigma \setminus D_\epsilon (Z)$ and 
$D_\epsilon(Z)$ is a disc of radius $\epsilon$ centred around $Z$. 
It is assumed that $h$ has an expansion of the form (\ref{hexpans}) around $Z$,
with the $\log$-term fixed but the coefficients $a,b,c,d$ etc. free to vary. 
The last two terms in (\ref{eq:TaubesAction}) can be seen as a contribution 
coming from a boundary action and are necessary to obtain a well 
defined variational principle, given that $h$ has the logarithmic
singularity $2\log|z-Z|$.
The $4 \log \epsilon$ term makes $S$ finite, and the variation of $4 a$ 
cancels the boundary term coming from the integration by parts. By 
requiring $S$ to be stationary under any variation 
$h\rightarrow h+\delta h$ with $\delta h$ vanishing at infinity and 
finite at the origin, one obtains Taubes' equation (\ref{Taubes1}). The delta 
function is absent since we have removed its support from the bulk 
action, but it is recovered by taking account of the logarithmic 
singularity. 

We now define the on-shell action to be the action $S$ evaluated for 
a solution of Taubes' equation. This depends on the vortex location
$Z$, and the background geometry. In section 2 we pointed out that the 
metric on the moduli space can be written in terms of the coefficient 
$b$ in (\ref{hexpans}). Here we show that, on-shell, $b$ is related to a 
derivative of $S$, and hence $S$ is part of the K\"ahler potential.

The derivative of $S$ with respect to $Z$ is
\begin{align}
\frac{\partial S}{\partial Z} = &\notag \lim_{\epsilon \to 0} 
\left\{ \frac{i}{2\pi} 
\int_{\Sigma_\epsilon} dz\wedge d\bar{z} \frac{\partial}{\partial Z} 
\left[2 \partial_z h\,\partial_{\bar{z}}h+\frac{\Omega}{\tau} 
\left(e^h-1-h\right) \right] \right.\\
&\label{SZ} \left. -\frac{i}{2\pi} \oint_{\gamma_\epsilon} d\bar{z} 
\left[2 \partial_z h\,\partial_{\bar{z}}h+\frac{\Omega}{\tau}
\left(e^h-1-h\right) \right]\right\} + 4 \frac{\partial a}{\partial Z} \,,
\end{align}
where the second integral comes from the variation with respect to $Z$ 
of the domain of integration $\Sigma_\epsilon$, giving an integral 
over its boundary $\gamma_\epsilon =\partial
\Sigma_\epsilon$. On-shell, we can make use of Taubes' equation 
to put the first integral in the form
\begin{align}
&\lim_{\epsilon \to 0} 
\left\{ \frac{i}{2\pi} \int_{\Sigma_\epsilon} dz\wedge d\bar{z} 
\left[ 2\partial_z\left(\frac{\partial h}{\partial Z}\,
\partial_{\bar{z}}h \right)
+2\partial_{\bar{z}}\left(\frac{\partial h}{\partial Z}\,
\partial_{z}h\right)\right] \right\} \nonumber \\
&=\lim_{\epsilon \to 0} 
\left\{ \frac{i}{\pi} \left( -\oint_{\gamma_\epsilon} d \bar{z}
\frac{\partial h}{\partial Z}\,\partial_{\bar{z}} h+\oint_{\gamma_\epsilon} dz 
\frac{\partial h}{\partial Z} \partial_z h \right)\right\} \nonumber \\
&=-4\frac{\partial a}{\partial Z} +3 \bar{b} \,,
\end{align}
where we have made use of the behaviour of $h$ near $Z$ and 
used a variant of Cauchy's residue theorem. For the second integral, 
in the limit $\epsilon\rightarrow 0$, the only term 
contributing is $\bar{b}/ (\bar{z}-\bar{Z})$ from 
$2 \partial_z h \partial_{\bar{z}} h$, so the integral is 
$-\bar{b}$, again using the residue theorem. Combining these integrals 
and the $4 \, \partial a /\partial Z$ term, we find that 
$\partial S / \partial Z=2\bar{b}$, and as $S$ is real,
\be
\frac{\partial S}{\partial \bar{Z}} = 2 b\,.
\ee

Using this result we can rewrite the conformal factor of the Samols 
metric (\ref{Sam1}) as
\begin{equation}
\tO(Z,\bar{Z}) = \Omega(Z,\bar{Z})+\tau\,\frac{\partial^2 S}
{\partial Z \partial \bar{Z}}\,.
\end{equation}
This means that the K\"ahler potential on the 1-vortex moduli space is 
given by
\begin{equation}
\tK (Z,\bar{Z})= \K (Z,\bar{Z}) + \tau \,S (Z,\bar{Z}) +\rm{const.}
\label{eq:Kahler}
\end{equation}
where $\K$ is the K\"ahler potential of $\Sigma$.

If we insist that $\K$ and $\tK$ have the same asymptotic form 
$ Z\bar{Z}$, and their difference tends to zero asymptotically, then $\tau S+\rm{const.}$ must tend to zero as the vortex 
location tends to infinity. For a vortex on an asymptotically planar 
surface, located far from the region where the Gaussian 
curvature $K$ and its derivatives differ significantly from zero, 
the profile function $h$ will be almost identical to that in the flat case;
only the exponential tail of $h$ will experience the region 
with significant $K$. Therefore 
\be
\lim_{ Z \to \infty} S(Z,\bar{Z}) = S_{flat}\,,
\ee
where $S_{flat}$ is the on-shell action $S$ evaluated on the solution 
$h_{flat}$ of the Taubes equation on the flat background. The constant 
term in (\ref{eq:Kahler}) must therefore be $-\tau S_{flat}$.

To facilitate the numerical studies it is better to simplify the 
on-shell action by making use of Taubes' equation again to rewrite
\begin{align}
2 \partial_z h \partial_{\bar{z}} h 
&=-2 h \partial_z\partial_{\bar{z}} h 
+ 2\partial_{\bar{z}} \left( h\partial_z h\right) \nonumber\\
&=-\frac{\Omega}{2 \tau}\,h\left( e^h-1\right) 
+ 2\partial_{\bar{z}} \left( h\partial_z h\right)\,.
\end{align}
This yields
\begin{equation}
S= \lim_{\epsilon \to 0} \left\{  \frac{i}{2\pi} \,
\int_{\Sigma_\epsilon} dz \wedge d\bar{z}\,\frac{\Omega}{\tau} 
\left[ \left( 1- \frac{h}{2}\right) e^h -\frac{h}{2} -1\right]
+\frac{i}{\pi} \oint_{\gamma_\epsilon} dz \, \left(h\,\partial_{z} h\right)
+4a+4 \log \epsilon\right\}\,.
\end{equation}
As $h\sim 2 \log \epsilon +a$ and $\partial_z h \sim \frac{1}{z-Z}$ 
for $\epsilon$ small, the residue theorem implies that
\begin{equation}
\frac{i}{\pi} \oint_{\gamma_\epsilon} dz \, \left(h\,\partial_{z} h\right)
=-4 \log \epsilon -2a\,.
\end{equation}
The $\log \epsilon$ terms now cancel and the remaining integral part of $S$ 
becomes well defined on  $\Sigma$ (the integrand still has a
logarithmic singularity at the vortex location but that is integrable) 
so that the limit $\epsilon \to 0$ can easily be taken.
We can simplify $S$ even further by noticing that
\begin{equation}
\frac{i}{2} \,\int_\Sigma dz \wedge d\bar{z}\,\frac{\Omega}{\tau} 
\left(1- e^h\right)=4\pi \,,
\end{equation}
since this integral is twice the magnetic flux. 
The on-shell action can therefore be written in the form
\begin{equation}
S=-\frac{i}{4\pi} \int_\Sigma dz\wedge d\bar{z} \,\frac{\Omega}{\tau}\,
h \left( 1+ e^h\right) +2a -4\,,
\label{Sonshell}
\end{equation}
which can be easily computed numerically once the solution to 
Taubes' equation is known. Equation (\ref{eq:Kahler}) takes the final form
\be
\tK = \K +\tau \left(S -S_{flat} \right)\,,
\ee
with $S$ given by (\ref{Sonshell}).

\end{document}